\documentclass{aa}
\usepackage[utf8]{inputenc}
\usepackage{color}
\usepackage{graphicx}
\usepackage{array}
\usepackage{breqn}
\usepackage{txfonts}
\usepackage{natbib}
\usepackage{booktabs,caption,fixltx2e}
\usepackage[flushleft]{threeparttable}
\bibpunct{(}{)}{;}{a}{}{,}

\begin{document}

\newcommand{\green}{\textcolor{green}}
\newcommand{\red}{\textcolor{red}}
\newcommand{\blue}{\textcolor{blue}}

\titlerunning{Galaxy-ICM interaction in clusters}
\authorrunning{Liyi Gu et al.}

\title{Implications of the mild gas motion\\
found with {\it Hitomi} in the core of the Perseus cluster
}
\author{Liyi Gu \inst{1,2}
\and
Kazuo Makishima \inst{1,3,4}
\and
Ryoji Matsumoto \inst{5}
\and
Kazuhiro Nakazawa \inst{6}
\and 
Kazuhiro Shimasaku \inst{7}
\and 
Naohisa Inada \inst{8}
\and 
Tadayuki Kodama \inst{9}
\and 
Haiguang Xu \inst{10,11,12}
\and 
Madoka Kawaharada \inst{13}
}

\institute{
RIKEN High Energy Astrophysics Laboratory, 2-1 Hirosawa, Wako, Saitama 351-0198, Japan
\and
SRON Netherlands Institute for Space Research, Sorbonnelaan 2, 3584 CA Utrecht, the Netherlands 
\and 
Department of Physics, Graduate School of Science, The University of Tokyo, 7-3-1 Hongo, Bunkyo-ku, Tokyo 113-0033, Japan
\and 
Kavli IPMU, The University of Tokyo, 5-1-1 Kashiwa-no-ha, Kashiwa, Chiba 277-8535
\and 
Department of Physics, Graduate School of Science, Chiba University, 1-33 Yayoi-cho, Inage-ku, 
Chiba 263-8522, Japan
\and 
Kobayashi-Maskawa Institute for the Origin of Particles and the Universe,
Nagoya University, Furo-cho, Chikusa-ku, Nagoya, Aichi 464-8601, Japan
\and 
Department of Astronomy, Graduate School of Science, The University of Tokyo, 7-3-1 Hongo, Bunkyo-ku, 
Tokyo 113-0033, Japan
\and 
Department of Physics, Nara National College of Technology, Yamatokohriyama, Nara 639-1080, Japan
\and 
Astronomical Institute, Tohoku University, Aramaki, Aoba-ku, Sendai 980-8578, Japan
\and 
School of Physics and Astronomy, Shanghai Jiao Tong University, 800 Dongchuan Road, Shanghai 200240, China
\and 
Tsung-Dao Lee Institute, Shanghai Jiao Tong University, 800 Dongchuan Road, Shanghai 200240, China
\and 
IFSA Collaborative Innovation Center, Shanghai Jiao Tong University, 800 Dongchuan Road, Shanghai 200240, China
\and 
Tsukuba Space Center (TKSC), Japan Aerospace Exploration Agency (JAXA), 2-1-1 Sengen, Tsukuba, 
Ibaraki 305-8505, Japan
}

\abstract{
Based mainly on X-ray observations,
studies are made on interactions
between the intra-cluster medium (ICM) in clusters of galaxies 
and their member galaxies.
Through (magneto)hydrodynamic and gravitational channels,
the moving galaxies are expected to drag the ICM around them,
and transfer to the ICM some fraction of 
their dynamical energies on cosmological time scales.
This hypothesis is in line with several observations,
including the possible cosmological infall of galaxies towards the cluster center,
found over redshifts of $z \sim 1$ to $z \sim 0$.
Further assuming that the energy lost by the galaxies
is first converted into ICM turbulence and then dissipated,
this picture can explain the subsonic and uniform ICM turbulence,
measured with {\it Hitomi} 
in the core region of the Perseus cluster.
The scenario may also explain several other 
unanswered problems regarding clusters of galaxies,
including what prevents the ICM from the expected radiative cooling,
how the various mass components in nearby clusters
have attained different radial distributions,
and how a thermal stability is realized 
between hot and cool ICM components
that co-exist around cD galaxies.
This view is also considered to pertain 
to the general scenario of galaxy evolution,
including their environmental effects.

}

\titlerunning{Galaxy-ICM interaction}
\authorrunning{L. Gu}
\maketitle

\section{Introduction}
\label{sec:intro}
The most dominant observed form of cosmic baryons 
is found as Intra-Cluster Medium (ICM), 
i.e., the X-ray emitting hot plasmas 
gravitationally confined in individual clusters of galaxies.
The ICM is still subject to a series of unsolved puzzles,
including the balance between cooling and heating.
Since the radiative cooling time of ICM at the center of rich clusters
is significantly shorter than the Hubble time $t_{\rm H}$,
these plasmas were thought to cool 
and gradually lose their pressure.
This will create inward plasma motion called cooling flows (CFs),
which further increase the density of the ICM in the central-region of the cluster,
and enhance its radiative cooling.
The operation of this feedback was apparently
supported by early imaging spectroscopic observations
of a number of clusters in soft X-rays.
Namely, clusters that host cD galaxies,
or cD clusters, ubiquitously show a series of 
characteristic  phenomena within $\sim 100$ kpc of the cD galaxy,
including an inward decrease in the ICM temperature,
and a strongly peaked brightness distribution therein \citep{fabian1994}.

As the X-ray imaging spectroscopy became available
in a broader energy band (up to 10 keV) with better energy resolution,
it was confirmed that the cluster core regions 
indeed host cooler X-ray emission,
hereafter called Central Cool Component (CCC).
However, essentially in all galaxy clusters and groups,
plasma components with the temperature 
below $\sim 0.5$ keV have been revealed to be much less 
than predicted by the CF scenario.
This puzzling discovery,
first made with {\it ASCA} \citep{ikebe1999, makishima2001}, 
was reinforced through observations with
{\it XMM-Newton} \citep{peterson2001, xu2002}, 
{\it Chandra} \citep{fabian2001}, and {\it Suzaku} \citep{gu2012}.
As a result, the nomenclature of ``cool-core clusters" 
was adopted instead of the ``CF clusters".
Apparently, the ICM is somehow heated against the radiative cooling, 
so that the expected CFs are suppressed or mitigated.

Among several candidate mechanisms proposed to explain the ICM heating,
the most popular one is the so-called AGN feedback scenario 
\citep{churazov2001, fabian2000, mcnamara2007, fabian2012}.
According to this conjecture,
the active galactic nucleus (AGN) of the central
(usually cD) galaxy in a cluster
operates in a kinetic/radio mode,
by accreting the surrounding cold/hot matter
at a sub-Eddington rate, 
and releasing most of the gravitational energy 
into radio jets or inflating bubbles instead of radiation.
The heating is considered to just balance the radiative ICM 
cooling through an inherent negative feedback mechanism;
an enhanced cooling will increase the ICM density 
surrounding the central black hole,
making the AGN more active, 
thus providing a higher heating luminosity 
to be deposited on the ICM.
Although the scenario is presented with many theoretical variants,
generally the highly-collimated AGN jets are assumed to be 
transporting the energy from the black hole to a very 
limited fraction of the ICM volume (i.e., not volume-filling).
Therefore, the scenario requires another mechanism 
to efficiently redistribute the deposited energy
over the cluster core volume.
One of the promising candidates 
was propagation of strong turbulence created by the AGN jets 
\citep{ensslin2006, chandran2007, scan2008, kunz2011}.
Then, the heat from the turbulence dissipation would spread over
the cool core within the cooling timescale, 
so as to offset the  CFs.

In its brief lifetime, 
the {\it Hitomi Observatory} \citep{takahashi2014}
obtained a set of X-ray spectra from 
central regions of the Perseus cluster of galaxies,
with the Soft X-ray Spectrometer (SXS)
which is a non-dispersive detector 
with a superb energy resolution.
Using the Fe-K line widths measured with the SXS,
the turbulence velocity dispersion
was found to be $\sim 164$ km s$^{-1}$ in a region
$30-60$~kpc from the central nucleus, and the shear in the bulk
motion to be $\sim 150$ km s$^{-1}$ 
across the central 60~kpc region \citep{hitomi2016}. 
A further analysis of the SXS data revealed 
that the velocity dispersion is relatively 
uniform within the central 100 kpc, 
except for the innermost core ($r < 20$~kpc) 
and the vicinity of an AGN cavity,
where mild increases were observed \citep{hitomi_tur}. 
The average energy density contained in the turbulence is 
estimated to be $\sim 4$\% of the total ICM thermal energy density.

These {\it Hitomi} results have brought some difficulties
into the ICM heating mechanisms 
that invoke the AGN activity and the consequent turbulence.
First, due to the low energy density, 
the turbulence would be able to sustain the X-ray emission 
only for a very short time, $8 \times 10^{7}$ years,
or $0.006$ $t_{\rm H}$ \citep{hitomi2016}.
Second, to maintain a stable balance 
between cooling and heating,
the turbulence must be replenished on the same timescale,
but this is not necessarily obvious. 
Furthermore, in $8 \times 10^{7}$ years,
such a low-speed ($\sim 164$ km s$^{-1}$) turbulence 
would propagate from the excitation sources
only over $\sim 13$~kpc 
which is far insufficient to heat 
the entire cool core \citep{fabian2017}.
In short, given the {\it Hitomi} results, 
it would be difficult to heat
the ICM in the cool core of the Perseus cluster
globally, and for a sufficiently long time,
if we consider only the random gas motions 
driven by the AGN jets and bubbles.

The above evaluations all assumed
that the X-ray emission from the cluster core,
contributing a large fraction ($\sim80\%$) 
of the total X-ray luminosity,
must be sustained by the turbulence energy
of the core-region ICM
which carries a small fraction ($\sim10\%$) of the overall ICM mass.
Therefore, the problem could be mitigated by 
assuming significant energy transport in other forms,
e.g., heat conduction from outer regions to the core \citep{voigt2004},
or sound waves from the AGN \citep{fabian2017, zweibel2018, bambic2018}.
Deep X-ray imaging of the Perseus cluster
hints for the latter, if the surface brightness fluctuations found in the core
are indeed sound waves from the bubbles \citep{fabian2003b, fabian2006, sanders2007}. 
What still remains controversial is
whether the sound waves can be dissipated
just as they propagate across the cool core region \citep{rus2004, fujita2005}.

Another possible way around these difficulties was proposed 
by \citet{lau2017} based on numerical simulations,
introducing a fair amount of changes to the AGN feedback model.
They assume a series of frequent small outbursts 
from the central AGN, instead of a few powerful ones. 
The induced turbulence field would then be low and flat,
because the input kinetic energy is low and steady.
However, they still need to call for 
accretion of member galaxies and subclusters 
to explain the mild line-of-sight velocity gradient 
observed with {\it Hitomi},
because such a gentle AGN feedback
would be too uniform to
explain the observed bulk velocity gradient.
A similar conclusion was reached by \citet{bourne2017} 
based on a different simulation.

Motivated by these difficulties,
in the present work we focus on
a relatively unexplored aspect of clusters of galaxies:
interactions between the two major baryonic components,
namely, the moving member galaxies and the ICM.
As briefly summarized in \S~\ref{sec:review},
this idea was first proposed by \citet[][ hereafter Paper I]{makishima2001}
based on {\it ASCA} observations 
and incorporating some semi-quantitative evaluations.
Since then, the view has been reinforced by a series of 
data-oriented studies (see \S~\ref{sec:supports})
with {\it XMM-Newton} \citep{takahashi2009,kawaharada2009},
{\it Chandra}, and {\it Suzaku} \citep{gu2012},
as well as by extensive X-ray versus optical comparisons
of clusters at various redshifts \citep{gu2013a, gu2016}.
The emerging scenario can potentially solve some of the 
issues with the ICM, including its temperature structures, 
the metal distributions in it,
and the suppression of CFs.

Our scenario also has rich implications for the ICM turbulence.
In fact, the assumed  physical interactions 
between the galaxies and ICM, 
and the consequent gravitational perturbations, 
will create magneto-hydro-dynamical (MHD)
turbulence in situ throughout the cluster core. 
The turbulence is naturally expected to be volume-filling,
thanks to the high galaxy density in the core region
(one per $50-100$ $\rm kpc^{3}$; \citealt{bud2012, gu2016}),
and must be subsonic
because the galaxies are moving through the ICM
with trans-sonic velocities.
Actually, a numerical simulation by \citet{ruszkowski2011}
based on a similar scenario predicts $\sigma=150-250$ km s$^{-1}$,
in a good agreement with the {\it Hitomi} measurements.
In the present paper, 
we discuss how this scenario has been reinforced by the {\it Hitomi} results.

\section{Review of the Proposed Scenario}
\label{sec:review}

As a preparation,
we briefly review, in this section,
our interpretations of the physics in central regions of cD clusters.
The first set of assumptions presented in \S~\ref{subsec:1st_set}
deal with spatial magnetic and temperature structures 
of the ICM in a static sense,
while the 2nd set in \S~\ref{subsec:2nd_set}
describe dynamics and cosmological evolution 
of the physical conditions therein.

\subsection{Basic assumptions (1): static aspects}
\label{subsec:1st_set}
In order to describe static magnetic structures 
of the ICM in central regions of cD clusters,
we start from a generally accepted view,
that the ICM  is hydrostatically confined by gravity,
and is magnetized to $\beta \sim 30 - 300$,
where $\beta$ is the thermal plasma pressure relative to the magnetic pressure.
That is, the magnetic energy density amounts to 
a fraction of percent to a few percent of 
the average thermal energy density  \citep{bohringer2016}.
In addition, we adopt the following first set of assumptions
(Paper I; \citealt{takahashi2009, gu2012}).
The overall configuration is referred to as ``cD corona" picture.
\begin{enumerate}
\item[S1]
The ICM in the cluster core region has 
relatively well-ordered magnetic structures,
because the cD galaxy is approximately 
standstill with respect to the ICM.
\smallskip
\item[S2]
The magnetic field lines (MFLs) in the cluster core region 
are classified into two types; 
closed loops with their both ends anchored to the cD galaxy,
and open ones with at most one end attached to it.
\smallskip
\item[S3]
The open-MFL regions are filled with the hot isothermal ICM,
whereas the closed MFLs confine a cooler and metal-richer plasma,
identified with the CCC,
to form a magnetosphere.
The two plasma phases are in an approximate pressure equilibrium.
\end{enumerate}

Of these assumptions, 
S1 is a natural consequence of the unique location of a cD galaxy,
i.e., the center of the gravitational potential.
The 2nd assumption is just a basic and general classification of MFLs,
as observed in solar coronae and interplanetary space.
A related discussion, on a possibility of
tangled MFLs, is presented in \S~\ref{subsubsec:4_velocities}.
The assumption S3 is also a prediction by plasma physics.
The open-MFL regions, connected to the outer ICM,
will be brought into a global isothermality on a time scale of $\sim 10^8$ yr,
by the high thermal conductivity along the MFLs.
In contrast, the CCC can
have a different temperature and metallicity,
because the particle/heat transport
is strongly suppressed across the MFLs.
The CCC plasma is considered to partially originate from the cD galaxy.

The cD corona model (S1-S3) may naturally explain 
several characteristic features seen in the cluster central regions. 
First, filamentary structures in the optical line emission
have been observed around many cD galaxies, 
on scales of $\sim 10-100$~kpc 
\citep{conselice2001, crawford2005, mcdonald2012, mcdonald2015}.
Considering that these structures are likely to trace the closed MFLs,
the assumed configuration is consistent with the observed  co-existence of 
different gaseous components in the core region,
some cool/dense while others hot/tenuous.

Second, the ICM around cD galaxies indeed shows 
stratified temperature structures as in S3.
The X-ray spectra from these regions require 
a two-phase (2P) modeling,
employing two temperatures which represent
the hot ICM and the CCC
\citep{ikebe1999, tamura2003, takahashi2009, gu2012, walker2015, dp2017}.
In contrast, those from outer regions can be described 
by a single-phase (1P) modeling.
The 2P condition cannot be an artifact 
arising from the projection effects,
because \citet{takahashi2009} and \citet{gu2012} showed,
through a careful deprojection procedure,
that the CCC and the hot phase spatially co-exist  
in the 3-dimensional core region  of cool-core clusters.
Since the hot phase is often $2-3$ times hotter than the CCC, 
the CCC must be confined by MFLs,
and thermally shielded from the hot phase.
The cool-phase gas is free to move along the closed MFLs,
so that it will be approximately in a pressure equilibrium with the hot phase.

Finally, the structure of intracluster MFLs (S1) can be directly 
probed through Faraday rotation measurements of radio signals 
from background sources or cluster member galaxies.
These measurements of the cluster central regions 
reveal coherence scales of $2-25$~kpc
\citep{taylor1993, feretti1999, govoni2001, eilek2002, clarke2004}, 
indicating that the MFLs may also
have a spatial order on such scales, instead of being completely tangled.

\subsection{Basic assumptions (2): dynamical aspects}
\label{subsec:2nd_set}
The 1st set of assumptions (S1-S3) above
are complemented by the following 2nd set of assumptions
(Paper I; \citealt{takahashi2009, gu2012, gu2013a, gu2016}),
which describe dynamical and evolutionary aspects of the problem.
\begin{enumerate}
\smallskip
\item[D1]
Non-cD member galaxies,
moving through the ICM at its trans-sonic velocities,
interact with the ICM, and transfer to it 
some fraction of their dynamical energies
on a cosmological time scale.
\smallskip
\item[D2]
The member galaxies, thus losing the energy,
gradually fall towards the center of the potential,
while supplying metal-enriched inter-stellar medium (ISM) to the ICM.
%
\smallskip
\item[D3]
The energy deposited onto the ICM will first take a form of MHD turbulence,
which will then be thermalized to provide 
a heating luminosity of $\sim 10^{44}$ erg s$^{-1}$
that is needed to balance the radiative cooling, 
particularly in the CCC.
\smallskip
\item[D4]
Throughout the heating process,
the two plasma phases are kept in a pressure equilibrium.
and are thermally stablized
by the Rosner-Tucker-Vaiana (RTV; \citealt{rosner1978}) mechanism
which is operating in solar coronae. 
\end{enumerate}

Of these assumptions,
probably the least accepted idea would be D1,
because galaxies are generally believed
to swim freely through the ICM 
with insignificant energy dissipation.
However, as explained in 
\S~\ref{subsec:support_d} and further in \S~\ref{subsubsec:interaction_gravity},
this popular belief is not necessarily warranted.
Furthermore, the assumption D2,
which is an immediate consequence of D1,
is now supported by indirect 
(\S~\ref{subsubsec:indirect}) 
and direct (\S~\ref{subsubsec:infall}) observations, 
as well as numerical studies.

We assign the entire \S~\ref{sec:turbulence} 
to the discussion on D3,
because it is most relevant to the {\it Hitomi} results.
Finally, D4 may appear too specific.
However, as described in \S~\ref{subsubsec:RTV},
it is a direct consequence of the cD corona configuration,
and is an essential ingredient of our overall scenario
because it  provides a very natural way to thermally stabilize the CCC.

\section{Scenario of Galaxy-ICM interaction}
\label{sec:supports}

Addressing the assumptions in the dynamical aspects
(\S~\ref{subsec:2nd_set}) is in essence
to find answers to the following three key questions:

\begin{itemize} \vspace*{-2mm}
\item[(1)]Do the ICM interact with the member galaxies,
and contribute significantly to their evolution?
%
%
\item[(2)] Is the dynamics of member galaxies
also affected appreciably by the presence of the ICM?
\item[(3)] Conversely, do the moving galaxies at all influence
 the thermodynamical evolution of the ICM?
\vspace*{-2mm}
\end{itemize}
Below, we examine these issues.

\subsection{Evidence for the interaction: ram pressure stripping and environmental effects}
\label{subsec:support_d}
First, we review observational evidence 
for the galaxy-ICM interaction,
and point out its important role in the evolution of member galaxies. 
These results will altogether answer
affirmatively the first key issue raised above.

As suggested in the early work by \citet{gunn1972}, 
and being confirmed by observations 
\citep[and references therein]{gorkom2004},
any gaseous interstellar material in a galaxy,
collectively called the ISM,
would feel the ram pressure of the ICM 
as the galaxy moves through the cluster. 
When the ram pressure exceeds the gravitational restoring force,
the ISM in outer galaxy disks will be stripped off, 
to form a tail in the trailing wake of the galaxy. 
While the loss of ISM will suppress the star formation 
in outer regions of the galaxy, 
the ram pressure will at the same time compress the ISM 
in central regions and the tails, to
enhance the star formation therein
\citep{fujita1999, gorkom2004, vollmer2006}. The efficient removal of the metal-rich ISM
is in line with the fact that the ICM contains a large amount of metals,
comparable to those contained in stars,
and the ICM metallicity is quite uniform up to the periphery
(described later).

Some statistical studies indicate that
the ram pressure effects are ubiquitous.
An H$\alpha$ and R-band imaging of 55 spiral galaxies in the Virgo cluster 
showed truncated star forming disks in 52\% of the sample galaxies, 
which are likely the product of galaxy-ICM interaction
\citep{koopmann1998, abadi1999, koopmann2004}. 
Although the so-called galaxy harassment \citep{moore1996}
could also be contributing, 
\citet{couch1998} used the {\it Hubble Telescope} data of three clusters 
to argue that the star formation is more efficiently truncated 
by the ram pressure than by harassment. The observed optical and \ion{H}{I} properties
motivated many authors (e.g., \citealt{bosch2013, rod2014})
to propose an evolutionary route of galaxies, 
beginning with field-like spirals and transforming, 
via long-term ram pressure stripping and passive fading,
into proto-S0 galaxies.

The statistical studies mentioned above can be generalized 
into the well known concept of ``environmental effects".
It is long known that, at low redshifts,
the galaxies in crowded environments predominantly have 
early-type morphology \citep{oemler1974, dressler1980}, 
red color \citep{pimbblet2002, balogh2004}, 
and suppressed star formation rates \citep{balogh1998, goto2003}. 
Although the main population of member galaxies 
in local clusters is thus red and inactive, 
the presence of blue star-forming galaxies in rich clusters 
at $z \sim 0.5$ has been known since the early work by \citet{butcher1984}. 
These two environmental effects, one spatial and the other temporal,
can be naturally and consistently explained by assuming
that the presence of the ICM and neighbouring galaxies
accelerates the galaxy evolution,
including the morphology/color changes
and the rapid decline in the star formation rate.

\subsection{Evidence of galaxy infall}
\label{subsec:support_gi}

We have so far answered affirmatively to the first key issue
presented at the beginning of this section;
the ICM indeed interacts with the moving galaxies,
and are likely to accelerates their evolution.
Next we address the key issue (2),
whether the interaction also affects 
the dynamics of galaxies, and cause them to slow down
and gradually fall towards the cluster center.
Below, direct observations of possible galaxy infall 
on cosmological timescales
(e.g., \citealt{gu2013a,gu2016}; \S~\ref{subsubsec:infall})
are augmented by several pieces of
indirect observational evidence from different aspects
(\S~\ref{subsubsec:indirect}).

\subsubsection{Radial distributions of baryons and metallicity therein}
\label{subsubsec:indirect}
In nearby clusters of galaxies,
the three major mass components are well known
to show systematically different radial profiles;
relative to dark matter (DM),
the ICM is slightly more extended,
whereas member galaxies are much more concentrated \citep{bahcall1999}.
Figure~\ref{fig:gnimr} (left) directly compares
the radial distributions of the galaxies and ICM,
both averaged over 119 clusters at $z<0.08$.
Within a radius of $\sim R_{500}$
beyond which the measurements become less accurate,
the ICM show a flatter radial distribution 
than the galaxy component. 
To quantify the difference, 
we introduce a quantity called
``galaxy light to ICM mass ratio" (GLIMR; \citealt{gu2013a}),
which is the ratio of the circularly (two-dimensionally) integrated 
galaxy light divided by the ICM mass integrated similarly.
Then, in nearby clusters, GLIMR typically decreases by a factor of 2
from $r= 0.2 \; R_{500}$ to $r= 0.5 \; R_{500}$ \citep{gu2013a, gu2016}.
One possible explanation to the GLIMR gradient 
would be provided by our assumptions D1 and D2,
which predict that galaxies should gradually sink to the center
by losing their dynamical energies to the environment,
and the energy deposited to the ICM will 
make it expand relative to the DM. 
If it works continuously,
the GLIMR profile will be steepened over time,
in a qualitative agreement with our scenario.

Since metals in the ICM must have originated from galaxies, 
and can serve as a tracer of the past galaxy distributions, 
their large-scale spatial behavior helps us 
to understand the present-day mass distributions.
Here, we should consider two important observational facts.
One is the result obtained with the {\it Suzaku} XIS,
that the ICM metallicity in outskirts 
(with the two-dimensional radius $r > 1$ Mpc) 
of several clusters are spatially quite constant 
at $\sim 0.3$ solar \citep{werner2013}. 
The observed constant metallicity at larger radii is 
usually taken for evidence
that the metal enrichment of ICM took place very early, 
at redshifts of $z =  2-3$ before the cluster formation,
via strong galaxy winds and/or AGN outflows 
\citep{fujita2008, werner2013}. 
Here, an implicit assumption is
that stars at these early epochs distributed out to very large radii,
and provided the initial metals to the ICM therein.
However, as indicated by the GLIMR behavior, 
the present-day clusters do not harbor, 
at their peripheries, the corresponding population of galaxies. 
Therefore, simply invoking the ``early metal enrichment'' view
alone would not easily bring into agreement 
the overall observational facts quoted so far.

As a natural way to bring the early ICM enrichment scenario 
into agreement with the present-day GLIMR profiles,
we assume that the metal-supplying stars actually 
formed galaxies in the peripheral regions, 
which then gradually fell towards smaller radii, 
as in our assumption D2. 
At the beginning of this process, 
the metals were transported to 
the intracluster space presumably via galactic winds \citep{fujita2008}, 
but as the galaxies moved to regions of higher ICM densities, 
e.g., $\lesssim 1$ Mpc, the role of ram-pressure stripping 
of metal-enriched ISM is considered to have become 
gradually more important \citep{fujita1999}.
This agrees with the very mild inward increase of the ICM metallicity,
observed on intermediate radii from 
$\lesssim 1$ Mpc down to $\sim 100$ kpc
\citep{matsushita2003, johnson2011, mernier2017}.

The other observational fact to be considered is the behavior 
of a quantity called ``metal mass to light ratio" (MMLR), 
i.e., the circularly integrated metal mass in the ICM 
divided by the galaxy light integrated in the same way. 
In many nearby clusters and groups, MMLR has been observed to increase, 
from the center to $\sim 100$ kpc, by nearly two orders of magnitude 
\citep{ksato2007, kawaharada2009, ksato2009b, ksato2009, sasaki2014}.
This outward MMLR increase must continue 
beyond 100 kpc out to larger radii, 
because we have MMLR $\propto$ metallicity/GLIMR, 
where the metallicity is rather constant to the periphery (the first fact),
and GLIMR decreases outwards (\S~\ref{subsubsec:infall}).
In short, the present-day galaxies are more concentrated 
relative to not only the ICM, but also the metals in it.
This can also be explained in a natural way by our assumption D2.


\subsubsection{Direct observations of galaxy infall}
\label{subsubsec:infall}

\begin{figure*}[!htbp]
\center
\resizebox{0.94\hsize}{!}{\includegraphics[angle=0]{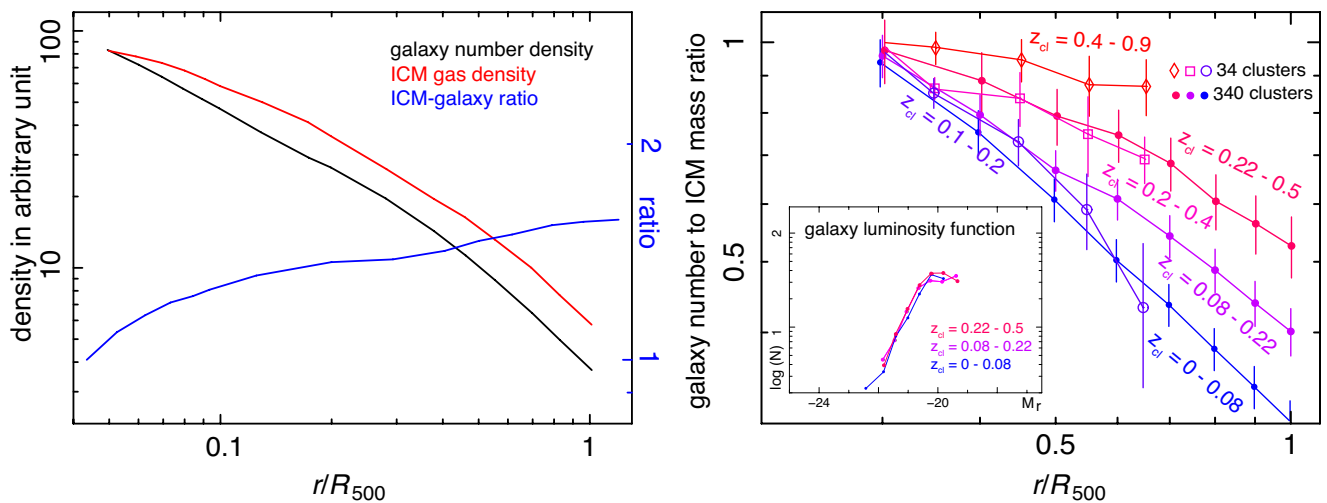}}
\caption{($left$) Projected radial density profiles 
(left ordinate) of the galaxy surface number (black) and the ICM (red) 
in nearby ($z \leq 0.08$) clusters,
normalized at a radius of 0.05$R_{500}$. 
Both data are taken from the sample of \citet{gu2016}. 
The ICM to galaxy density ratio is plotted in blue
(right ordinate).
($right$) Observed GNIMR profiles 
averaged over clusters in different redshift bins.
This plot shows a combination of two samples, 34 clusters ($z=0.1-0.9$) 
from \citet[open symbols]{gu2013a} and 340 clusters ($z=0-0.5$) from \citet[filled symbols]{gu2016}.
Error bars are at the 1$\sigma$ level.
The $r$-band luminosity functions of the identified member galaxies from
the 340 cluster sample in different redshift bins are shown in the inset.
}
\label{fig:gnimr}
\end{figure*}

Although the spatial distributions of the different 
components in nearby clusters have been shown
(\S~\ref{subsubsec:indirect})
to give support to our dynamical assumptions,
a more direct confirmation of the scenario was awaiting for
a comparison of distant clusters with their nearby counterparts.
This has finally been carried out successfully by \citet{gu2013a};
the authors studied a carefully selected sample of
34 clusters in a redshift range of $z=0.1-0.9$,
both in the optical and X-ray frequencies,
to compare the optical and X-ray angular extents
of individual clusters in the sample. 
All the selected clusters have relaxed X-ray morphology,
and harbor cD galaxies at their centers.
Since the utilized optical data were photometric rather than spectroscopic,
the galaxy membership in each cluster was determined
using photometric redshifts, supplemented by an offset pointing.
Then, as reproduced in Fig.~\ref{fig:gnimr} (right)
with open symbols,
the GLIMR profiles of these clusters were 
found to be originally flat at $z \sim 0.9$,
and were evolving to become steeper as the redshift decreases.
We believe that this provides one of the fundamental 
evolutionary effects ever observed from clusters of galaxies.

Utilizing a much larger sample consisting of 340 SDSS clusters,
though with a shallower redshift coverage of $z=0.0-0.5$,
\citet{gu2016} carried out a more sensitive
follow-up study of the galaxy-to-ICM evolution.
This time, a fraction of the optical data were spectroscopic,
and a quantity ``galaxy number to ISM mass ratio" (GNIMR)
was used instead of GLIMR.
Then, as shown in Fig.~\ref{fig:gnimr} (right)
with filled symbols,
the X-ray vs. optical comparison again 
revealed a clear relative evolution,
which agrees quantitatively with the first result by \citet{gu2013a}. 
In addition, \citet{gu2016} found
that the galaxies became radially more
concentrated with respect to the DM component as well. 
At the same time, the luminosity functions of 
the member galaxies used by \citet{gu2016} were confirmed 
approximately the same over the relevant redshift range
(inset to Fig.~\ref{fig:gnimr} right).
Therefore, the enhancement of galaxies 
in the cluster center towards $z \sim 0$
cannot be ascribed to a recent galaxy/star formation.
Instead, the galaxies from the cluster periphery 
likely have moved inwards, 
with respect to both the ICM and DM, 
on a timescale of $\geq 6$ Gyr.

The dynamical evolution of member galaxies in clusters has 
also been investigated using numerical simulations,
where the galaxy-scale processes such as ram pressure stripping 
have been reproduced utilizing 
progressively finer spatial grids ($\sim$kpc or less)
A recent hydrodynamical work by \citet{armitage2018} explored 
the evolution of velocity dispersion of member galaxies 
as a function of time spent inside the cluster. 
As shown in their Figure~8, the member galaxies are indeed found 
to be slowing down after their first passage through the cluster core. 
If normalized to the velocity of dark matter, 
the velocities of member galaxies decrease 
by $\sim 20$\% after spending 10~Gyr in the cluster. 
Naturally these galaxies will fall towards the cluster center 
as they become slower than the ambient dark matter particles.
Although these authors mainly considered gravitational dynamical friction,
the slowdown of their galaxies was found little dependent on the total galaxy mass,
contrary to what is expected from the
pure gravitational interaction (Eq.~\ref{eq:Fg}).
Therefore, we infer that other forms of galaxy-cluster interaction, 
such as the ram pressure process, are likely to be 
contributing to the slowdown of their model galaxies.
Similar galaxy slowdown and infall have been observed in 
other hydrodynamical simulations (e.g., \citealt{ye2017}).

\subsection{Evaluation of energetics}
\label{subsubsec:interaction_general}

In this section, we address the key issue (3);
how the suggested galaxy slowdown/infall
will affect the energetics of the ICM. 
Since observational constraints are rather limited. 
we instead perform some analytic estimations,
focusing on cluster-averaged energy transfer 
associated with the galaxy infall
(\S~\ref{subsubsec:infall_energy}),
the energy releases expected from different physical channels 
(\S~\ref{subsubsec:interaction_energy}),
and the balance between the ICM cooling at the cluster core 
and the proposed form of heating (\S~\ref{subsubsec:core_heating}).

\subsubsection{Energy release associated with the observed infall}
\label{subsubsec:infall_energy}

Let us evaluate the energy release associated with the galaxy infall.
In a cluster with a galaxy velocity dispersion of $v=1000$ km s$^{-1}$,
a typical galaxy with a total mass of 
$\sim 1\times 10^{11}$~$M_{\odot}$ 
will lose a dynamical energy of $\sim 3 \times 10^{60}$~erg
while falling from a radius of 
$R_{500}$ ($\sim 1$ Mpc) to the center.
Therefore, in a cluster with $\sim 100$
galaxies \citep{vdb2018} thus falling to the core region,
the energy released in this way throughout 
the process is estimated to be 
\begin{equation}
E_{\rm gal} \sim 3 \times 10^{62} 
\left( \frac{M_{\rm gal}}{10^{13}M_\odot} \right)~{\rm erg}
\label{eq:Egal}
\end{equation}
where $M_{\rm gal}$ is the total mass of the moving 
galaxies. Further considering that this process takes place
on a time scale $\tau_1$ 
which is comparable to the Hubble time 
as suggested by \citet{gu2013a,gu2016}, namely
\begin{equation}
\tau_1 \sim t_{\rm H} = 4.5 \times 10^{17}~{\rm s}~,
\label{eq:tau1}
\end{equation}
the time-averaged energy-release rate
by the galaxies would be 
\begin{equation}
L_{\rm infall} = E_{\rm gal}/\tau_1 
\sim  7\times 10^{44}~{\rm erg~s}^{-1}~.
\label{eq:infall_luminosity1}
\end{equation}
%

A more reliable and  conservative estimate can be obtained 
by calculating the gravitational energy change 
between the radial distributions of member galaxies, 
observed in high-redshift and present-day clusters. 
Specifically, using the average galaxy light profiles 
of the two subsamples in \citet{gu2013a}, one
at $z=0.45-0.90$ (nine clusters with a mean redshift of 0.653)
and the other at $z=0.11-0.22$ (nine clusters with a mean of 0.174),
and assuming a luminosity-dependent mass-to-light ratio
\citep{cappellari2006} for the galaxies,
this dynamical energy change in a typical cluster becomes
\begin{equation}
\Delta E_{\rm gal} = 6 \times 10^{61} \left(\frac{M_{\rm cl}}{2\times10^{14} M_{\odot}}\right) ~~{\rm erg}~
\label{eq:infall_energy}
\end{equation}
where $M_{\rm cl}$ is the total cluster mass.
Note that the mass-to-light ratio reported in \citet{cappellari2006} 
refers to the galaxy central region; 
$\Delta E_{\rm gal}$ would increase significantly
if considering the observed outward increase 
in the mass-to-light ratio of individual galaxies (e.g., \citealt{fukazawa2006}).
Even when we ignore this factor and retain the above
conservative estimate for the mass-to-light ratio,
this $\Delta E_{\rm gal}$ is a reasonable fraction of 
$E_{\rm gal}$ in Eq.(\ref{eq:Egal}),
considering that the relevant time lapse is 
$\sim 1.2 \times 10^{17}$ s = $0.27 t_{\rm H}$.
As a result, the energy release rate by the galaxies 
in a typical cluster over this time interval becomes
\begin{equation}
\label{eq:infall_luminosity2}
L_{\rm infall} \sim 5 \times 10^{44} 
\left(\frac{M_{\rm cl}}{2\times10^{14} M_\odot}\right) 
~~{\rm erg~s^{-1}}
\end{equation}
in a reasonable agreement with Eq.(\ref{eq:infall_luminosity1}).

\begin{table}[!htbp]
\centering
\caption{Properties of the hot and cool components in the centers of Perseus cluster, Centaurus cluster, and Abell~1795, within the specified radius.}
\label{tab:chlum}
\begin{threeparttable}
\begin{tabular}{ccccccccccc}
\hline \hline
Objects  & Perseus $^{a}$& Centaurus $^{b}$& Abell 1795$^c$\\
\hline
Radius (kpc)    &  60     &  150      &    144\\
Cool $kT$ (keV) &$2.9-3.6$ & $1.7-2.0$ & $2.0-2.5$\\
Hot $kT$ (keV)  &$4.5-4.7$ &  3.8      & $4.7-6.3$ \\
Cool $L_{\rm X}$$^{d}$&  4.1  &  0.1   & 1.4 \\
Hot $L_{\rm X}$$^{d}$ &  2.2  &  0.3  & 2.5  \\
\hline
\end{tabular}
\begin{tablenotes}
\item[$(a)$] \citet{hitomi_temp}.
\item[$(b)$] \citet{takahashi2009}.
\item[$(c)$] \citet{gu2012}.
\item[$(d)$] Over 0.3--10 keV, in units of $10^{44}$ erg s$^{-1}$.
\end{tablenotes}
\end{threeparttable}
\label{tbl:core_luminosities}
\end{table}

Table~\ref{tab:chlum} summarizes
0.3--10 keV luminosities of the two components,
observed from core regions of three representative clusters.
Thus, $L_{\rm infall}$ in Eq.~(\ref{eq:infall_luminosity1}) and
Eq.~(\ref{eq:infall_luminosity2}) is high enough
to balance the radiative output of CCC observed from typical clusters.
For reference, Table~\ref{tab:chlum} implies that
a considerable fraction of the X-ray luminosity 
from the core region
is still carried by the hot component.
As already pointed out in Paper I,
this is because the centrally peaked surface brightness
is observed around a cD galaxy not only in the cool component,
but also in the hot component,
reflecting a hierarchical structure in the 
gravitational potential in the core region of
such clusters \citep{xu1998,ikebe1999,takahashi2009,gu2012}.
Consequently, the cool-component luminosity 
was often overestimated previously (Paper I).

\subsubsection{Energy transfer through galaxy-ICM interactions}
\label{subsubsec:interaction_energy}


We have so far shown 
that the overall energy release from the infalling galaxies
is sufficient to sustain the CCC luminosity.
Although this {\it integrated} evaluation gives 
a partial answer to the key issue (3),
we still need to investigate detailed {\it differential} energetics,
to understand how a moving galaxy transfers its energy to the environment,
in particular to the ICM. As described below, 
we can think of a few different interaction modes.

The most straightforward interaction mode is
the head-wind ram pressure (\S~\ref{subsec:support_d}).
The overall ram-pressure force exerted to each galaxy 
from the inflowing ICM can be written as 
(equation 5.28 in \citealt{sarazin1988})
\begin{equation}
\label{eq:rp}
F_{\rm RP} = \pi R_{\rm int}^2 \rho v^2,
\end{equation} 
where $R_{\rm int}$ is 
the effective hydrodynamic/MHD
interaction radius of a moving galaxy, 
$\rho$ is the ICM density, and $v$ is the galaxy velocity. 
Although this force works in the first place
on the gaseous components of each galaxy, 
it will indirectly slow down the entire galaxy,
including the stellar and DM components. 
This is because a gaseous component, 
 initially bound to the galaxy by gravity,
will gravitationally pulls back the entire galaxy,
if it is displaced/stripped by the ram pressure
(details in \S~\ref{subsubsec:interaction_gravity}).


Beside the ram pressure, various transport processes 
at the boundary between the ICM and a moving galaxy, 
might also have a similar effect on the galaxy. 
In particular, the ICM,
when regarded as a viscous flow, 
would exert friction on a galaxy, as
\begin{equation}
\label{eq:vis}
F_{\rm VIS} = \pi R_{\rm int}^2 \rho v^2 12/R_{\rm e},
\end{equation}
where $R_{\rm e}$ is the Reynolds number
of the ICM. This $F_{\rm VIS}$ 
becomes significant only when the ambient
ICM has a sufficiently large viscosity \citep{nulsen1982}.
However, this might not be actually the case; 
as reported recently in \citet{irina2019}, 
the ICM viscosity in the Coma cluster  
is likely $< 0.1$ times the Spitzer value. 
Converting the reduced viscosity into the Reynolds number
as $R_{\rm e} > 500$
through Eq.(2) of \cite{brunetti2007},
$F_{\rm VIS}$ would become only a few percent of $F_{\rm RP}$.


Although the aerodynamic effects considered above
requires the presence of ICM,
the orbiting galaxies also interact with the host cluster 
in a gravitational mode \citep{balbus1990}
even without the ICM. 
This is so-called galaxy-cluster dynamical friction,
which arises when a moving galaxy 
gravitationally scatters DM particles,
to slightly change the DM distribution around it.
Using a time-evolving perturbation theory, 
\citet{ostriker1999} derived the dynamical drag force 
which operates on a member galaxy with a mass $m_{\rm gal}$,
moving with a transonic motion, as
\begin{equation}
\label{eq:Fg}
F_{\rm G} = 4 \pi \rho_{\rm tot} (G\; m_{\rm gal})^2 / v^2,
\end{equation}
where $\rho_{\rm tot}$ is the total mass density dominated by DM.
Thus, the effect is more significant for
massive (and hence mostly elliptical) galaxies, 
which carry most of the stellar mass, 
rather than the more abundant low-mass galaxies. 

Employing realistic spatial distributions of
$\rho$ and $\rho_{\rm tot}$, 
and assuming $R_{\rm int} \sim 10$ kpc,
\citet{gu2013a} calculated how the motions of model galaxies
in a typical cluster potential are affected by the drag force
$F_{\rm RP} + F_{\rm G}$.
It was confirmed that their orbits indeed decay significantly
on a time scale of the order of Eq.(\ref{eq:tau1}),
even though the results naturally depend
on the mass and initial positions of the galaxies;
e.g., $F_{\rm RP}$ and $F_{\rm G}$ are dominant 
for galaxies with $m_{\rm gal}=10^{10}~M_\odot$ 
and $m_{\rm gal}=10^{11}~M_\odot$, respectively.
To quickly check the effect of $F_{\rm RP}$,
it may suffice to calculate as 
\begin{dmath}
\label{eq:tau1_again}
\tau_1 \sim \left( \frac{3}{2} m_{\rm gal} v^2 \right) / (F_{\rm RP}\: v)
= 1.2 \: t_{\rm H}\times \ 
\left(\frac{m_{\rm gal}}{1 \times 10^{11}M_\odot} \right)
\left(\frac{R_{\rm int}}{10{\rm kpc}} \right)^{-2}
\left(\frac{n_{\rm e}}{10^{-3}{\rm cm}^{-3}} \right)^{-1}
\left(\frac{v}{10^3 {\rm km/s}} \right)^{-1}
\end{dmath}
where the normalizing electron density,
$n_{\rm e}=10^{-3}$ cm$^{-3}$, 
represents a typical value
averaged within $\sim 500$ kpc.
Therefore, the moving galaxies will lose their 
dynamical energies approximately 
on a time scale of $\sim t_{\rm H}$,
when the effect is averaged over a cluster.
Because of the $n_{\rm e}^{-1}$ dependence,
the time scale is expected to gets shorter,
to $\sim 0.1 t_{\rm H}$ at the cluster core region.

In place of Eq.(\ref{eq:tau1_again}),
we may more directly conduct an order-of-magnitude 
estimate of the heating luminosity by the ram pressure.
We then obtain from Eq.(\ref{eq:rp})
\begin{dmath} 
\label{eq:RP_haeting}
L_{\rm RP} = N F_{\rm RP} \/ v = N \pi R_{\rm int}^2 \rho v^3\\
 \sim 6 \times 10^{44} \left( \frac{N}{100} \right)
   \left( \frac{R_{\rm int}}{10~{\rm kpc}} \right)^{2}
   \left(\frac{n_{\rm e}} {10^{-3}~{\rm cm^{-3}}} \right)
   \left(\frac{v}{10^3~{\rm km/s}} \right)^{3}
   ~{\rm erg~s^{-1}}
\end{dmath}
where $N$ is the total number of in-falling galaxies.
Of course, this estimate is essentially equivalent
to the combination of Eq.(\ref{eq:infall_luminosity1})
and Eq.(\ref{eq:tau1_again}).
In addition, Eq.(\ref{eq:RP_haeting})
reconfirms the more crude estimate made 
in \S~\ref{subsubsec:infall_energy} and \citet{gu2013}.


\subsubsection{How to heat the cluster core}
\label{subsubsec:core_heating}

\begin{figure*}[!htbp]
\center
\resizebox{0.9\hsize}{!}{\includegraphics[angle=0]{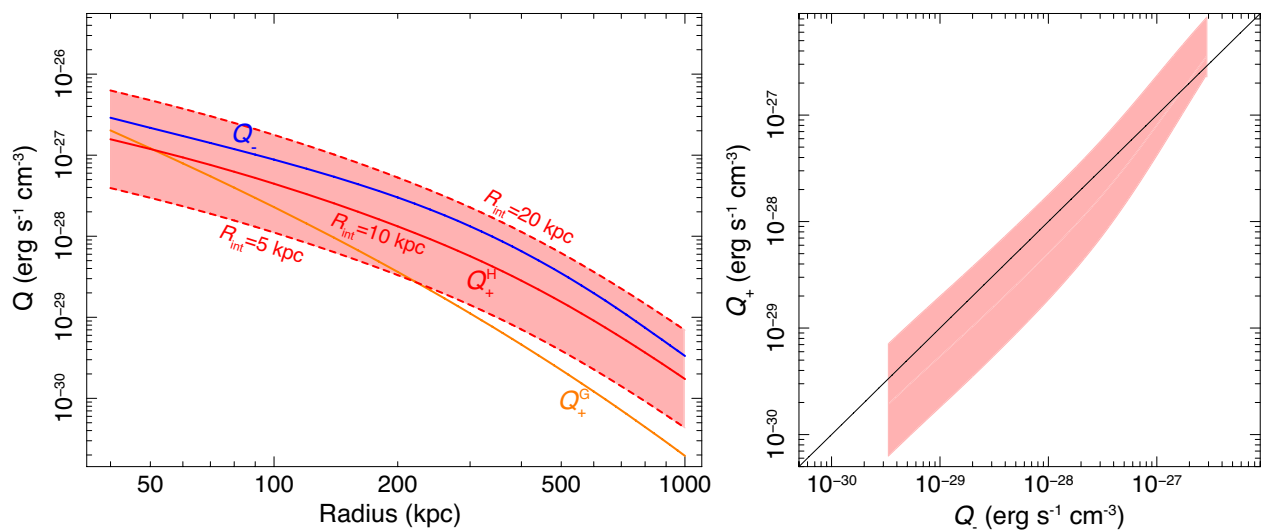}}
\caption{
(left) Radial profiles of the volume cooling rate $Q_{-}$ (blue)
and the volume heating rate $Q_{+}$ from
hydrodynamical (red; $Q_{+}^{\rm H}$) 
and g-mode (orange;$Q_{+}^{\rm G}$) interactions,
shown as a function of 3D radius.
The shadow region shows the range caused by 
different hydrodynamical interaction radii, 
$R_{\rm int}$ in Eq.(\ref{eq:rp}). 
(right) a comparison between $Q_{-}$ and $Q_{+}=Q_{+}^{\rm H}+Q_{+}^{\rm G}$,
over various 3D radii.
The vertical extent of the shadow region reflects 
the uncertainty of $R_{\rm int}$,
like in the left panel. 
Solid line indicates the exact balance between $Q_{-}$ and $Q_{+}$.
}
\label{fig:cool-heat}
\end{figure*}

So far, our evaluation of energetics has been 
based mostly on cluster-averaged estimates.
However, this is obviously insufficient,
because the ICM density, 
and hence the radiative cooling time too,
varies by a few orders of magnitude from the center 
to the periphery of each cluster,
and the essential problem is how to deposit 
sufficient heating luminosity in the core region
where the ICM most vitally needs to be heated.

To address this important issue,
we compare the cooling and heating rates of ICM,
considering their dependence on the 3D radius $R$.
The ICM cooling rate per unit volume
is expressed as 
\begin{equation} \label{eq:Q-}
Q_{-}= \Lambda(T,Z) \; n_{\rm e} n_{\rm i}  
\sim \Lambda(T,Z) \; n_{\rm e}^2
\end{equation}
where $n_{\rm i}$ is the ion  density,
and $\Lambda(T,Z)$ is the plasma cooling function 
depending on the temperature $T$ and metallicity $Z$.
On the other hand, Eq. (\ref{eq:rp}) means 
an ICM volume heating rate as
\begin{equation} \label{eq:Q+}
Q_{+}= \pi R_{\rm int}^2 \; n_{\rm gal} \;  \rho v^3
\end{equation}
where $n_{\rm gal}$ is the local galaxy number density.
For reference, volume integration of this equation
reduces to Eq.(\ref{eq:RP_haeting}).
Then, assuming $R_{\rm int}^2$ to be constant
and using $n_{\rm e} \propto \rho$,
we obtain
\begin{equation} \label{eq:Q+/Q-}
\frac{Q_{+}}{Q_{-}}  
\propto  \frac{n_{\rm gal} \rho v^3}{\Lambda(T,Z)\; n_{\rm e}^2}
\propto  \left( \frac{n_{\rm gal}}{n_{\rm e}}\right)
         \frac{v^3}{\Lambda(T,Z)} ~ .
\end{equation}
%

Let us examine this $Q_{+}/Q_{-}$ ratio
for its $R$-dependence.
First, both $n_{\rm gal}$ and $n_{\rm e}$ depend strongly on $R$,
but their gradients mostly cancel out,
and leave us with a factor $\leq 2$ higher $n_{\rm gal}/n_{\rm e}$
ratio at the center than in the periphery, 
as indicated by Fig. 2 (left).
Next, $\Lambda(T,Z)$ depends only weakly on $R$,
and is at most 30 percent higher at the center,
when representing the core region with
$T_{\rm c}=2 $ keV and 1 Solar metallicity,
whereas the outre region with
$T_{\rm h}=5 $ keV and 0.3 Solar.
Finally, the radial profiles of $v$ 
have long been studied optically
(e.g., \citealt{zhang2011}; \citealt{bilton2018}).
The results vary;
a central increase of $v$ is observed from some clusters,
whereas a decrease from others.
In cool-core clusters which are of our main interest,
$v$ generally increases weakly towards the center
\citep{bilton2018}.
Considering all these factors,
we infer that $Q_{+}/Q_{-}$ is
approximately flat across the cluster,
within a factor of a few.

The above inference needs two remarks.
One is the projection effect,
as the spherical velocity profile of $v$ is obviously
different from its projected profile to be observed.
However, this effect is considered to be minor
like in the deprojection analysis of the ICM temperature.
The other is the expected radial decrease of $R_{\rm int}$,
due to the inward decrease of 
the fraction of gas-richer galaxies.
This effect can be taken into account
by replacing $n_{\rm gal}$ in Eq.(\ref{eq:Q+/Q-})
with the number density of blue galaxies,
$n_{\rm gal}^{\rm B}$.
According to \citet{barkhouse2009}, 
the ratio $n_{\rm gal}^{\rm B}/n_{\rm gal}$
becomes halved at the center,
but a 20\% increase towards the center is observed 
in the ratio $v^{\rm B}/v$ \citep{bilton2018},
where $v^{\rm B}$ is the velocity dispersion of blue galaxies.
Thus, $Q_{+}/Q_{-}$ would still be relatively flat
within a factor of a few.

To verify the above estimates, 
we quantitatively computed $Q_{+}$ and $Q_{-}$
based on actual observations.
That is, $Q_{-}$ was calculated as a function of $R$,
using the radial profiles of the ICM density,
temperature, and the metallicity,
all averaged over the 340 clusters 
with $z=0.0-0.5$ from \citet{gu2016}.
Similarly, in calculating $Q_{+}$,
its hydrodynamical contribution, denoted as $Q_{+}^{\rm H}$,
was evaluated using the sample-average profiles 
of $n_{\rm gal}$ and $\rho$ also from \citet{gu2016},
and the mean $v$ profile for a sample of 10 cool-core clusters 
in \citet{bilton2018} (their Fig. 4, right panel). 
Although $Q_{+}^{\rm H}$ consists of the ram pressure (Eq.~\ref{eq:rp})
and the viscous friction (Eq.~\ref{eq:vis}) terms,
the latter was neglected as argued in \S~\ref{subsubsec:interaction_general}.
The largest uncertainty in $Q_{+}^{\rm H}$
is the value of $R_{\rm int}$,
which should depend on the types of galaxies, 
and possibly on stages of the in-falling process as well.
In Fig.~\ref{fig:cool-heat} (left), 
we therefore calculated $Q_{+}^{\rm H}$
for three typical cases, 
$R_{\rm int} = 5$ kpc, 10 kpc, and 20 kpc.
Thus, for a fixed value of $R_{\rm int}$, 
the $Q_{+}^{\rm H}$ and $Q_{-}$ profiles
indeed depend very similarly on $R$
from the cluster center to $\sim 1$ Mpc,
and the $Q_{+}^{\rm H}/Q_{-}$ ratio is close to unity
if $R_{\rm int}$ is in between 10 and 20 kpc.
Here we have assumed
that all the dynamical energy losses from galaxies 
are in the end thermalized; 
if this is not the case, the $Q_{+}^{\rm H}$ profile 
needs to be scaled by the thermalization efficiency.

The g-mode interaction (Eq.\ref{eq:Fg}) is 
essential for a complete view of the heating. 
Although this process will convert the dynamical energy
of moving galaxies primarily into those of DM,
the produced fluctuations (both in time and position)
of the local gravitational potential
would ultimately be spent in the ICM heating
(via sloshing and similar effects).
As the g-mode strength depends critically on the galaxy mass, 
we estimated it separately for two galaxy subsets, 
low-mass galaxies with $m_{\rm gal} \leq 10^{11} M_{\odot}$, 
and the remaining high-mass ones. 
For each subset, we calculated Eq.(\ref{eq:Fg}),
again using the mass-sorted average $n_{\rm gal}$
profiles from \citet{gu2016}
and the $v$ profiles from \citet{bilton2018}. 
The $\rho_{\rm tot}$ information, also taken from \citet{gu2016}, 
is common to the two subsets. 
As shown in Fig.~\ref{fig:cool-heat} (left),
the heating rate $Q_{+}^{\rm G}$ 
from the gravitational interaction,
summed over the two subsets,
has a more centrally-peaked profile than $Q_{+}^{\rm H}$,
because of several effects, 
e.g., the more concentrated profile of 
$\rho_{\rm tot}$ than $\rho$,
and a stronger central clustering of massive galaxies.
In short, the g-mode interaction may contribute significantly 
to the heating at the core region, 
assuming efficient dissipation of
DM fluctuations on the ICM,
but much less relevant outside it.

In Fig.~\ref{fig:cool-heat} (right), 
we directly compare, at each $R$,
$Q_{-}$ against the total heating rate,
$Q_{+}=Q_{+}^{\rm H}+Q_{+}^{\rm G}$.
Although the uncertainties from $R_{\rm int}$ are again large, 
we confirm an approximate balance between the two quantities.
Thus, the galaxy-driven heating scenario has been shown 
to be matched very well with the $R$-dependent ICM cooling rate,
if $R_{\rm int}$ takes a value of $\sim 10$ kpc
which is actually suggested by the observed 
cosmological galaxy in-fall (\S 3.2.3; \citealt{gu2016}).

The quantitative verification in Fig.~\ref{fig:cool-heat}
may still leave us with one basic question.
The ICM thermal energy in the core region is 
far insufficient to sustain the X-ray emission
against its short cooling time;
indeed, this motivated the cooling-flow hypothesis.
Then, how would the sustained heating be available by the
galaxies in the core region,
of which the dynamical energy content must be
at most comparable to the ICM thermal energy therein?
The answer can be obtained if we notice
that an appreciable faction of these galaxies
must have larger orbits, 
and happened to be near the center at present,
rather than localized therein.
As already considered in the orbital calculation by \cite{gu2013a},
such a galaxy will repeatedly cross the core 
with a typical interval of $\sim 0.1 t_{\rm H}$,
and lose, in every crossing, 
a small fraction ($\sim 10\%$) of its dynamical energy.
Thus, the heating energy can be supplied by member galaxies 
which are distributed over a much larger volume
than the cluster core;
the energy budget discussed so far
in \S~\ref{subsubsec:infall} 
and \S~\ref{subsubsec:interaction_general}
should be understood in this context.

The essential difference between the galaxies' 
dynamical energy and the ICM thermal energy is 
that the former can be transported by the ballistic motions 
of galaxies on a time scale of $\sim 0.1 \/ t_{\rm H}$,
whereas the latter must be transported via electron conduction
which would take $\sim t_{\rm H}$ for $l \sim 1$ Mpc.
In other words, galaxies are cooled {\it globally},
because a considerable fraction of the overall galaxies
participate in the core heating,
whereas the ICM would be cooled (if no heating) very {\it locally}.

\section{ICM turbulence}
\label{sec:turbulence}

\subsection{Galaxy motion and ICM turbulence}
\label{subsec:gal_turbulence}

\subsubsection{An overview}
\label{subsubsec:turb_overview}
As shown in \S~\ref{sec:intro} and by \citet{hitomi_tur}, 
the ICM in the Perseus cluster core was revealed by {\it Hitomi} 
to be amazingly quiet.
The observed spatial field of the ICM velocity dispersion 
is roughly composed of two components, 
a flat field of $\sim 100$~km s$^{-1}$ 
spreading on a $\sim 100$~kpc scale, 
and regions of enhancement to $\sim 200$~km s$^{-1}$ 
in the innermost 20~kpc and at an AGN-inflated bubble. 
Although the direct AGN feedback might be responsible for the latter,
the origin of the former is still unclear.
Therefore, other possibilities are being explored.
For instance, \citet{zuhone2018} propose that
the turbulence originates from sloshing motion of the Perseus cool core,
in response to gravitational perturbations by in-falling subclusters.
In this model, the gas motion is fully explained 
by the energy input from subcluster mergers.
A similar scenario was proposed by \citet{inoue2014}.


In this section, 
we employ our assumptions D1-D3
to explore a somewhat different idea
that the ICM turbulence is mainly produced by the galaxy motion.
Obviously, the effects of moving galaxies are less energetic
than the mergers,
but much more frequent and ubiquitous.
Then, we try to answer the core question of the paper: 
is this scenario capable of explaining the 
essence of the  {\it Hitomi} measurement?

The dynamical energy of moving galaxies will be 
transferred to the ICM entropy at least through three channels.
One is direct ICM heating via, e.g., 
anomalous Joule dissipation of galaxy-induced electric currents.
Another is a channel wherein the galaxies' energy 
is first converted 
to kinetic energies of local ICM motions,
which are then dissipated as ICM entropy.
The other is an electromagnetic channel,
such as amplification of magnetic fields
and excitation of Alfv\'enic waves;
these energies will be 
transmitted to individual particles,
through, e.g., magnetic reconnection and wave damping.
Below, we consider only the 2nd channel,
which is thought to be dominant
\citep{norman1999,dennis2005}.

The local ICM motions induced by galaxies
that are moving with transonic velocities
will take a form of, e.g., sound waves,
vorticity around individual galaxies, 
and wakes trailing behind them \citep{vi2017},
that is, ICM {\it turbulence}.
The energy injection to the turbulence would 
take place on spatial scales of galaxies,
e.g., through a process known as resonant drag instability
\citep{seligman2019} in which plasma perturbations 
are enhanced when the galaxy's velocity matches
the phase velocity of propagating perturbations 
(e.g., sound waves in this case).
The turbulence will then spread through the ICM, 
and split down to smaller scales via the turbulence cascades.
Although the scenario is thus simple,
details of turbulence in a compressible and magnetized 
plasma are rather complicated.
Therefore, below we discuss only general aspects of the problem.

\subsubsection{Characterization of the observed turbulence}
\label{subsubsec:4_velocities}
Let us begin with comparing
four characteristic velocities involved here.
The first one is the sound velocity $s$ in the ICM,
which takes a value as
\begin{equation}
s = 
 \left(\frac{2 \gamma kT}{m_{\rm p}} \right)^{1/2}
= 1.2 \times 10^3 \left(\frac{T}{5~{\rm keV}} \right)^{1/2}~ {\rm km}~{\rm s}^{-1} 
\label{eq:sound_speed}
\end{equation}
where $\gamma=5/3$ is the adiabatic index,
and the temperature of $T=5$ keV 
applies to the Perseus cluster (the hot component).
For simplicity, the plasma is approximated as hydrogenic.
Next is the velocity dispersion of the galaxies,
denoted as $v$ as before, 
which is comparable to $s$ as widely recognized.

The third one is the Alfv\'{e}n velocity, given as
\begin{equation}
v_{\rm A} = \frac{B}{(4 \pi \rho)^{1/2} }
= 1.0 \times 10^2 
\left( \frac{B}{5~\mu G} \right) 
\left( \frac{n_{\rm e}}{10^{-2}~{\rm cm}^{^3}} \right) ~ {\rm km}~{\rm s}^{-1} 
\label{eq:Alfven_speed}
\end{equation}
where the magnetic-field intensity $B=5~\mu$G
and the ICM density $n_{\rm e}=10^{-2}$ cm$^{-3}$ used for normalization
are both appropriate in the Perseus core region.
The plasma beta, i.e., the thermal gas pressure 
relative to the magnetic pressure, becomes
\begin{equation}
 \beta = (2/\gamma) (s/v_{\rm A})^2 \sim 170~.
 \label{eq:beta}
\end{equation}
The last item is just the ICM turbulence velocity dispersion,
$\sigma \sim 160$ km s$^{-1}$ as measured with {\it Hitomi}.
Therefore, we find
\begin{equation}
 s \sim v \gg v_{\rm A} \sim \sigma.
\label{eq:4_velocities} 
\end{equation}
In short, the galaxy motion is transonic but super-Alfv\'{e}nic,
whereas the ICM turbulence is subsonic but trans-Alfv\'{e}nic.
It is  noteworthy
that the measured $\sigma$ is close
to the calculated $v_{\rm A}$.

Because $v \sim s$ holds,
the moving galaxies are not expected 
to create strong shocks in the ICM.
Because of $v \gg v_{\rm A}$,
they will not produce, either, so-called slow shocks
(discontinuities in the slow-mode longitudinal waves),
to be realized with sub-Alfv\'{e}nic plasma flows.
In the regime of Eq.(\ref{eq:4_velocities}), instead,
each moving galaxy will bend MFLs, and create kinks in them,
where thin sheets of electric currents should be formed.
These current sheets are expected to provide promising sites 
where the turbulence and electromagnetic energies
are efficiently dissipated,
e.g., via magnetic reconnection and Joule heating.


Under the condition of Eq.(\ref{eq:4_velocities}),
the turbulence will propagate through the ICM with two modes;
the longitudinal sound waves, 
and the transverse shear (or torsional) Alfv\'en waves.
Since $s \gg v_{\rm A}$, the magnetosonic velocity
$\left(s^2 + v_{\rm A}^2 \right)^{-1/2}$
is essentially the same as $s$.
Although MFLs are perturbed by these fluctuations,
we do not expect strong entanglement of them,
because the turbulence is trans-Alfv\'{e}nic.
Therefore, the ordered MFL structure assumed in S1
should remain valid.

\subsubsection{The expected turbulence velocity dispersion}
\label{subsubsec:sigma_estimates}
As described in \S~\ref{subsubsec:turb_overview},
we assume 
that the energy $\Delta E_{\rm gal}$ of Eq.~(\ref{eq:infall_energy})
will mostly flow through the channel of ICM turbulence
before dissipated into entropy,
and that this provides the dominant heating source for the ICM.
Then, the energy temporarily stored at present in the form of
ICM turbulence must be a certain fraction of $\Delta E_{\rm gal}$.
As a result, the ICM turbulence velocity dispersion 
$\sigma$ will satisfy
$\frac{3}{2}M_{\rm ICM} \sigma^2 < \Delta E_{\rm gal}$,
where $M_{\rm ICM}$ is the total ICM mass.
We hence obtain from Eq.~(\ref{eq:infall_energy}),
\begin{equation}
\label{eq:sigma_estimate1}
\sigma < 320 \: \left(\frac{M_{\rm ICM}}{10^{13} M_{\odot}}\right)^{-0.5} 
\left(\frac{M_{\rm cl}}{10^{14} M_{\odot}}\right)^{0.5}
{\rm km~s}^{-1}~.
\end{equation}
In short, the turbulence due to the galaxy motion 
must be subsonic in a rich cluster,
in agreement with the {\it Hitomi} results.

The above argument can be formulated 
in a more generalized context.
Suppose that the energy of galaxies, Eq.(\ref{eq:Egal}),
is transferred, on a time scale of $\tau_1$
(Eq.\ref{eq:tau1}, Eq.\ref{eq:tau1_again}),
to the ICM turbulence energy,
which in turn is dissipated into entropy \citep{irina2014} 
on another time scale $\tau_2$.
If the overall energy flow is in an approximate steady state,
we should expect
\begin{equation}
M_{\rm ICM} \sigma^2 \sim M_{\rm gal} v^2 \times (\tau_2/\tau_1)~.
\end{equation}
This yields an estimate, averaged over each cluster, as
\begin{equation}
\sigma = v \sqrt { \left( \frac{M_{\rm gal}}{M_{\rm ICM}} \right)
      \left( \frac{\tau_2}{\tau_1} \right) } 
       \sim 550 \:\left(\tau_2/\tau_1 \right)^{1/2}~{\rm km~s}^{-1},
\label{eq:sigma_estimate2}
\end{equation}
where we employed $v \sim 1000$ km s$^{-1}$
after Eq.~(\ref{eq:sound_speed}),
and $M_{\rm gal} /M_{\rm ICM} \sim 0.3$
which is appropriate for a cluster-averaged estimate.
This is consistent with Eq.(\ref{eq:sigma_estimate1}), 
if $\tau_2/\tau_1 \lesssim 0.3.$

For a further estimate focusing on the core region,
let us assume
that the turbulence is dissipated on its crossing timescale, 
namely, $\tau_2 \sim l_{\rm tur} / \sigma$, 
where $l_{\rm tur} $ is the largest driving scale 
(injection scale) of the turbulence,
which may be taken to be the length scale of galaxies, 
$l_{\rm gal} \sim 20$~kpc \citep{irina2014, irina2018}.
The general consistency between dissipation and crossing timescales 
for compressible and subsonic MHD turbulence has been validated
by numerical simulations as reported in, e.g., \citet{haugen2004}.  
Substituting this into Eq.(\ref{eq:sigma_estimate2}),
and solving it for $\sigma$, we obtain
\begin{dmath}
\sigma \sim ( M_{\rm gal}/M_{\rm ICM} )_{\rm c}^{1/3} \;
v^{2/3} \; (l_{\rm gal}/\tau_1 )^{1/3} 
\sim 195 \left( \frac{M_{\rm gal}}{M_{\rm ICM}} \right)_{\rm c}^{1/3}
\left( \frac{v}{10^3} \right)^{2/3} 
\left( \frac{l_{\rm gal}}{20{\rm kpc}} \right)^{1/3}
\left( \frac{\tau_1}{0.1\/ t_{\rm H}} \right)^{1/3}~~{\rm km\;s}^{-1}
\label{eq:sigma_estimate3}
\end{dmath}
where we utilized the estimate $\tau_1 \sim 0.1 t_{\rm H}$,
which applies to the cluster core
as noticed just after Eq.(\ref{eq:tau1_again}).
The first factor, $(M_{\rm gal}/M_{\rm ICM})_{\rm c}$,
is the galaxy-to-ICM mass ratio in the core region,
which may be taken as $\sim 1$. 
Thus, the predicted turbulence velocity
is fully consistent with the {\it Hitomi} measurement,
considering various uncertainties involved here.

From these values of $\sigma$, we obtain
\begin{equation}
   \tau_2 \sim l_{\rm gal}/\sigma \sim 7 \times 10^{-3} \; t_{\rm H}
 \label{eq:tau2}
\end{equation}
in the core region.
Therefore, the ICM turbulence therein
is inferred to dissipate on a time scale
comparable to the cooling time 
at the very center, $\sim 6 \times 10^{-3} t_{\rm H}$.
As a corollary to Eq.(\ref{eq:tau2}), we find
\begin{equation}
  \tau_2/\tau_1 \sim 0.07 
 \label{eq:tau2/tau1}
\end{equation}
which means that about 10\% of 
the dynamical energies of galaxies in the core region 
is stored temporarily as the ICM turbulence energy.
Although this is rather small 
as evident from the {\it Hitomi} results,
it can sustain the ICM entropy against the radiative cooling
because it is continuously replenished by a large number of galaxies,
some of which have larger orbits and 
happened to be in the core region.


\subsubsection{Spatial uniformity of turbulence}
\label{subsubsec:turbulence_uniformity}
At this stage, the remaining task is to examine
whether the proposed scenario can also
explain the observed uniformity of $\sigma$,
on spatial scales from $\sim 20$ kpc to $\sim 100$ kpc.
We may first consider the uniformity on 
rather small spatial scales as $<20$ kpc.
The ICM turbulence created by a single galaxy would occupy 
only a limited cluster volume localized around its trajectory.
However, in a massive cluster with numerous galaxies 
crossing the core region,
their loci, each with a typical width of 20~kpc \citep{roediger2015}, 
will fill up nearly the entire core volume 
in $\sim 10^8$ yr, the timescale on which the turbulence is
dissipated according to Eq.(\ref{eq:tau2}).
Therefore, when averaged for $\gg 10^8$ yr,
the ICM turbulence would be rather uniform 
within the cluster core region \citep{sub2006}.

We next consider how $\sigma$ would depend on $R$ 
on relatively large scales as $>50$ kpc. 
For this purpose,
let us break up the argument in \S~\ref{subsubsec:core_heating},
and consider the behavior of $\sigma$ as 
a quantity intervening between the heating and cooling.
Supposing that $Q_+$ and $Q_{-}$ are approximately balanced 
as in Fig.~\ref{fig:cool-heat},
the steady-state condition of the turbulence is then described,
from Eq.(\ref{eq:Q+}), as
\begin{equation}
\pi R_{\rm int}^2 \; n_{\rm gal} \;  \rho v^3 
= Q_{+} = Q_{-}
= \rho \sigma^2 /\tau_2 
\label{eq:flow_balance}
\end{equation}
which yields
$\sigma^2 = \pi R_{\rm int}^2 n_{\rm gal} v^3 \tau_2$
as $\rho$ cancels out.
Further approximating both $R_{\rm int}$ and $v$ 
as spatially constant, we obtain
\begin{equation}
\sigma^2 \propto n_{\rm gal} \; \tau_2  ~.
\label{eq:sigma_scalingH}
\end{equation}
Therefore, the problem reduces to
the behavior of $\tau_2$,

If the fluid were incompressible, 
$\tau_2$ would depend 
neither on the mean fluid density,
nor the turbulence strength.
However, its behavior becomes different in the ICM
which is clearly compressible.
According to a theoretical study by \citet{yoshizawa1997},
the turbulence dissipation in a compressible and
low-Mach number fluid is a nonlinear phenomenon,
so that $\tau_2$ gets shorter as $\sigma$ increases.
For simplicity, we may describe 
this effect as $\tau_2 \propto \sigma^{-k}$,
using an empirical index $k>0$.
Combining this with Eq.~(\ref{eq:sigma_scalingH}),
we obtain
\begin{equation}
    \sigma \propto \left( n_{\rm gal} \right) ^{1/(k+2)}~.
   \label{eq:Yoshizawa2}
\end{equation}
If assuming $k \sim 1$ (following Eq.~\ref{eq:sigma_estimate3}), 
$\sigma$ would vary by only a factor of $\sim 2$
over a spatial scale of several hundred kpc, 
across which $n_{\rm gal}$ changes by
an order of magnitude as in Fig.~\ref{fig:gnimr} (left).

In addition to the MHD processes as above, 
the in-falling massive galaxies can create the ICM turbulence 
also by exciting gravity (g-mode) waves 
\citep{balbus1990, ruszkowski2011}
as in Eq.(\ref{eq:Fg}).
In this case, the galaxies will pull, through dynamical friction, 
the local DM and ICM away from their original positions,
and their restoration by gravity and sound waves
creates a low-frequency oscillation in the ICM. 
As the galaxies gradually fall towards the cluster center,
the resulting g-mode waves would also 
propagate inward, 
and fill the cluster core volume.
Through the same argument as above
but using Eq.(\ref{eq:Fg}) instead of Eq.(\ref{eq:rp}),
we obtain, in place of Eq.~(\ref{eq:sigma_scalingH}),
\begin{equation}
\sigma^2 \propto n_{\rm gal} \; (\rho_{\rm tot}/\rho) \;\tau_2
\label{eq:sigma_scalingG}
\end{equation}
where $m_{\rm gal}$ and $v$ were approximated as constant.
Therefore, corresponding to Eq.(\ref{eq:Yoshizawa2}),
we obtain 
\begin{equation}
    \sigma \propto \left[ n_{\rm gal} \; (\rho_{\rm tot}/\rho) \right]^{1/(k+2)}~.
    \label{eq:Yoshizawa3}
\end{equation}
As the ratio $\rho_{\rm tot}/\rho$ depends only mildly on $R$,
the expected behavior of $\sigma$ is again 
similar to the case of Eq.(\ref{eq:Yoshizawa2}).

In summary, our scenario can explain,
not only the value of $\sigma$,
but to some extent
the observed uniformity of $\sigma$ as well.

\medskip
\subsection{Numerical simulations}
\label{subsec:simulations}

Among many numerical simulations of astrophysics
of clusters of galaxies,
the effects of magnetic fields on the galaxy vs. ICM interaction
has been addressed by only several,
including \citet{asai2004, asai2006, asai2007}, \citet{dursi2008}, and \citet{suzuki2013}. 
In the MHD framework that takes into account
radiative cooling and anisotropic heat conduction, 
these authors studied interactions between an uniformly inflowing ICM,
and a denser plasma gravitationally confined in a simulated galaxy.
This numerical setup is similar to the condition
considered in \S \ref{subsubsec:interaction_general},
although their model galaxy is a kind of intermediate
between a cD galaxy and a moving member in our scenario.
They have found several essential features,
including excitation of MHD turbulence on spatial scales of the galaxy,
formation of loop-shaped cooler regions around it, 
and a suppression of radiative cooling.
These results are in support of our scenario.

The g-mode interaction between the member galaxies 
and the ICM was studied 
with a three-dimensional simulation by \citet{ruszkowski2011}. 
Although they treated the ICM in an MHD scheme,
they ignored (magneto-) hydrodynamical interactions 
between the ICM and galaxies. 
The gas motion in the simulated clusters was found 
to have a velocity of  $100-200$ km s$^{-1}$, 
in good agreement with the {\it Hitomi} measurements.
The cluster-average gas motion 
approximately scaled as $R^{-1/2}$, 
which is qualitatively similar to Eq.(\ref{eq:Yoshizawa3})
assuming $n_{\rm gal} \sim R^{-(k+2)/2}$.

Based on the above simulation,
\citet{ruszkowski2011} calculated the heating luminosity
of the cool core, in a similar way to those 
in \S \ref{subsubsec:sigma_estimates}, as 
\begin{dmath}
L^{\rm cc}  \sim \rho \sigma^{3} V_{\rm cc}/ l_{\rm tur}
= 2 \times 10^{43} 
\left(\frac{n_{\rm e}}{10^{-2}\,{\rm cm^{-3}}} \right) 
\left(\frac{\sigma}{150 \, {\rm km/s}}\right)^{3} 
\times \left(\frac{R_{\rm cc}}{100 \,{\rm kpc}}\right)^{3} 
\left(\frac{l_{\rm tur}}{150 \, {\rm kpc}}\right)^{-1} 
{\rm erg~s^{-1}},
\label{eq:diss}
\end{dmath}
where $V_{\rm cc}$ and $R_{\rm cc}$ are 
the volume and radius of the cool core, respectively,
and $l_{\rm tur}\sim 150$ kpc is the typical length scale
of turbulence they found in their simulations.
This $l_{\rm tur}$ is considerably longer 
than the dissipation length, $l_{\rm gal} \sim 20$ kpc,
which we have assumed so far.
Consequently, $L^{\rm cc}$ fell 
below the cooling luminosity of the CCC of rich clusters
(Table~\ref{tbl:core_luminosities}).
If, however, the MHD interactions between the galaxies 
and ICM are properly taken into account,
$l_{\rm tur}$ could become much shorter,
e.g., down to $l_{\rm tur} \sim l_{\rm gal}$
as we have assumed, 
and hence $L^{\rm cc}$ could increase.

In spite of the insufficient heating by $L^{\rm cc}$,
the runaway cooling was interestingly suppressed 
in the simulation of \citet{ruszkowski2011}.
The authors discuss a possibility of 
heat inflow from outer to inner regions,
via heat conduction and/or turbulent diffusion
\citep{ruszkowski2011}.
Therefore, it is suggested that the ICM cooling
in the core region is at least partially offset
by such inward heat transport from outer regions.

More recently, the combined MHD and g-model interactions 
between the ICM and moving galaxies were studied numerically 
with an adaptive-mesh MHD approach by \citet{vi2017}. 
Their results (particularly movies) clearly reveal 
that each galaxy strongly interacts with the ICM,
deposits cooler ISM onto the intra-cluster volume,
and produce ICM turbulence on spatial scales of galaxies or less.
Most of the clusters they simulated are mildly turbulent 
($\sigma = 50-300$ km s$^{-1}$) within the central $\sim 500$ kpc.
In addition, the gas motion is volume filling,
and reasonably isotropic on large scales. 
Thus, the work by \citet{vi2017} can successfully explain
the {\it Hitomi} measurements in terms of the 
the galaxy-ICM interaction,
and provide a convincing support to our viewpoint.
This work also suggests that the galaxy-ICM interaction 
will systematically amplify the cluster magnetic fields, 
and drive magnetic evolution on cosmological timescales.

\subsection{Other observations of ICM turbulence}
\label{subsec:turb_observations}

Except the {\it Hitomi} results,
successful observations of the ICM turbulence are so far still rare, 
because of the obvious instrumental limitations.
Actually, the reports are  limited to a few observations 
using line broadening \citep{sanders2010, pinto2015},
resonant scattering \citep{ogo2017}, 
and surface brightness fluctuations \citep{irina2014}, 
in central regions of brightest clusters of galaxies. 
These different approaches have well converged to a subsonic 
gas motion with $\sigma=100-300$ km s$^{-1}$.
With the current data, however,
it is not possible to isolate gas motions
driven by the galaxy motion,
from those due to other excitation mechanisms
such as buoyant motions from the feedback of the central AGN. 
A full understanding of the galaxy-ICM interaction needs to wait 
for dedicated spatially-resolved observations
with future high-resolution X-ray spectrometers.

Despite the paucity of direct evidence, 
there are already a few hints showing turbulence driven by galaxies.
Optical/radio spectroscopic observations 
detected significant Doppler broadening in emission lines
from the cool gas tails produced by ram pressure stripping,
and the tail turbulence was actually measured in a few objects 
\citep{vollmer2006, mcdonald2012}. 
For instance, analyzing \ion{H}{I} clouds
which are ram-pressure-removed 
from  galaxies in the Virgo cluster, 
\cite{abramson2011} derived 
the velocity dispersions of the clouds as a few tens of km s$^{-1}$ 
to $\sim 150$ km s$^{-1}$ \citep{kenney2004}, 
in broad agreement with the {\it Hitomi} measurement in Perseus. 
However, these should be treated as a consistency check, 
rather than direct evidence for our scenario, 
as the physical link between the tail turbulence 
and the ICM turbulence is still unclear.

Recently, \citet{eckert2017b} estimated 
the ICM turbulence in a sample of 51 clusters,
indirectly through ICM density fluctuations,
and found that the power of their radio halo
depends strongly on the turbulence as
$P_{\rm radio} \propto \sigma^{3.3}$.
This scaling suggests a possible relation 
between the turbulent motions in the ICM
and the population of accelerated particles.
Then, by extrapolating the scaling
to low-$P_{\rm radio}$ regimes, 
we can estimate $\sigma$ on sub-cluster scales,
assuming that the underlying physics 
should not be strongly scale dependent.
Based on an Effelsberg 1.4~GHz observation, 
\citet{vollmer2004} detected an extended radio halo 
of $P_{\rm radio} \sim 1 \times 10^{22}$ W Hz$^{-1}$ 
centered on the giant elliptical M86 in the Virgo cluster, 
and attributed the results to
on-going interactions between M86 and the Virgo ICM. 
Then, the above scaling predicts 
the radio halo to have $\sigma \sim 100$ km s$^{-1}$. 
A similar extended radio halo is found centered on NGC~4839, 
a group of galaxies falling into the Coma cluster
and emitting a 1.4~GHz radio power of 
$8 \times 10^{22}$ W Hz$^{-1}$ \citep{deiss1997}.
the scaling relation predicts $\sigma \sim 180$ km s$^{-1}$. 
These estimates are close to the {\it Hitomi} measurement.

To summarize, our scenario,
that the mild ICM turbulence
results from the galaxy-ICM interactions,
has been supported from several 
aspects, including the ability to explain the 
value of $\sigma$ (\S~\ref{subsec:gal_turbulence}),
a few numerical simulations (\S~\ref{subsec:simulations}), 
and some indirect observations (\S~\ref{subsec:turb_observations}).


\section{Discussion}
\label{sec:discuss}

\begin{table*}[!htbp]
\centering
\caption{Evidence for the proposed scenario.}
\label{tab:evidence}
\begin{threeparttable}
\begin{tabular}{cccccccccccc}
\hline
Evidence                     & Source$^{*}$             & Assumptions$^{\dagger}$   &Section \\
\hline
Ordered structures of the central magnetic fields           & O   & S1/S2    & \S~\ref{subsec:1st_set}\\
Co-existence of discrete plasma phases around  cD galaxies  & O   & S1/S2/S3/D4& \S~\ref{subsec:1st_set}\\
Long-trailing ram pressure tails behind galaxies           & O/N & D2      & \S~\ref{subsubsec:interaction_general}\\
Metal transport from galaxies and the MMLR profiles         & O   & D1/D2   & \S~\ref{subsubsec:indirect}\\
Radial distributions of the 3 components in nearby clusters & O   & D1/D2   &\S~\ref{subsubsec:indirect}\\
In-fall of member galaxies from $z=0.9$ to $z=0$            & O   & D1/D2   &\S~\ref{subsubsec:infall}\\
A detailed balance between $Q_{-}$ and $Q_{+}$              & C   & D1/D3 & \S~\ref{subsubsec:core_heating}\\
Mild turbulence in the ICM                                  & O/N & D1/D3    & \S~\ref{subsubsec:sigma_estimates},
                                                                              \S~\ref{subsec:simulations}\\
Spatial uniformity of the ICM turbulence                    & O/N & D3      & \S~\ref{subsubsec:turbulence_uniformity},
                                                     \S~\ref{subsec:simulations}\\
\hline
\end{tabular}
$^{*}$: O=observation, N=Numerical simulation, C=Calculation/estimation\\
$^{\dagger}$: See \S~\ref{subsec:1st_set} and \S~\ref{subsec:2nd_set} for the definition.
\end{threeparttable}
\end{table*}

Previous studies of clusters of galaxies,
either observational, theoretical, or numerical,
treated the galaxies and the ICM almost independently,
and interplay between them was considered 
only in rather limited aspects
such as local stripping of galactic diffuse media.
The present study, in contrast, has directly considered 
interactions between the high-density/low-entropy galaxies, 
and the low-density/high-entropy ICM,
i.e., the two major baryonic components of clusters.
Thus, adopting the two sets of simple assumptions
(\S\ref{subsec:1st_set} and \S\ref{subsec:2nd_set}),
and considering the (magneto-) hydrodynamic and g-mode
effects (\S\ref{subsubsec:interaction_general}),
we have developed a view
that a large amount of energy actually flows 
from galaxies to the ICM,
and that this phenomenon ubiquitously affects 
dynamics, energetics, and evolution of clusters.
As summarized in Table~\ref{tab:evidence},
an accumulating pieces of evidence give 
a support to our overall scenario.
Among them, of particular importance are 
the evidence of the cosmological galaxy in-fall
(\S~\ref{subsubsec:infall}),
and the consistency with the  
{\it Hitomi} results (\S~\ref{sec:turbulence}).
Our picture, however, has still two missing or inadequate pieces: 
(1) how do moving galaxies actually receive significant 
drag force from their host cluster,
and (2) what maintains the thermal stability 
between the hot ICM and the cool core.
Below, we address these two questions,
with (1) being a direct continuation 
from \S~\ref{subsubsec:interaction_energy}.

\subsection{Force transmission via gravity}
\label{subsubsec:interaction_gravity}

Although we have so far argued 
that moving galaxies experience a variety of ICM vs. ISM interactions,
a vital question still remains;
how would the inflowing ICM exerts drag force
on the stellar and DM components
which dominate the mass of each galaxy?
For this purpose, we need to consider the role of gravity.

The most common form of gravitational interaction 
is  the dynamical friction between each galaxy
and the host cluster (Eq.\ref{eq:Fg}). 
If it is fully responsible for the possible GLIMR or GNIMR evolution
(\S~\ref{subsubsec:infall})
reported in \citet{gu2013a, gu2016}, 
we would naturally  expect a mass-dependent radial 
segregation of member galaxies  \citep{nath2008}. 
In addition, galaxies in high-$\rho_{\rm tot}$ environments 
would fall faster than those in low-$\rho_{\rm tot}$ systems. 
However, \citet{gu2013a, gu2016} found 
that the GNIMR (or GLIMR) profiles are not much different
between brighter and fainter galaxy subsamples, 
or between lower-mass and higher-mass cluster subsamples.
Strictly speaking, lower-mass galaxies are observed 
to fall to the cluster center somewhat less rapidly
than the average galaxies \citep{gu2013a,gu2016},
but the difference is limited.
It is therefore difficult to attribute the observed 
galaxy in-fall solely to the dynamical friction.
This is consistent with the results presented 
in Fig.~\ref{fig:cool-heat}.

Here, we introduce a new recipe for the ICM vs. galaxy interaction,
combining the hydrodynamical and gravitational effects.
As shown by the numerical work of \citet{roediger2015},
the ram pressure by the inflowing ICM will
first strip a galaxy of its ISM,
outside a radius where the ram pressure balances 
the gravitational restoration force \citep{gunn1972}.
Then, within that radius and toward downstream of the galaxy,
a long-lasting tail will be formed 
by the ISM which survived the stripping.
According to current observations (\S\ref{subsec:support_d})
and numerical simulations \citep{roediger2015},
this displaced ISM tail (including new stellar populations born therein)
is massive enough to gravitationally pull the whole 
stellar and DM components of the galaxy toward downstream,
until the tail is fully stripped. 
In this way, the ICM can indirectly exert drag force
to the non-gaseous components of each moving galaxy.

Employing a simple analytic approach,
let us quantitatively examine 
whether the above idea actually works or not.
The ISM tail may be approximated
by a sphere of radius $R_{\rm ism}$
with a uniform mass density $\rho_{\rm ism}$,
which is displaced just by the same $R_{\rm ism}$
from the potential center of the galaxy
which has a mass of $m_{\rm g}$.
Then, the gravitational restoration force
working on the sphere will be
$G\/ m_{\rm gal} \left(\frac{4\pi}{3} 
\rho_{\rm ism} R_{\rm ism}^3 \right)/ R_{\rm ism}^2
= \left(\frac{4\pi}{3}\right) G \/ m_{\rm gal} \rho_{\rm ism} R_{\rm ism}$.
Equating this with the ram pressure $\pi R_{\rm ism}^2 \: \rho v^2$
working on the ISM sphere,
where $\rho$ is the ICM density as before, 
the critical radius for stripping is obtained as 
\begin{equation}
 R_{\rm ism} =  \frac{4}{3} \left( \frac{G \/ m_{\rm gal}}{v^2} \right)
  \left(\frac{\rho_{\rm ism}}{\rho} \right).
\label{eq:R_ism}
\end{equation}
This is close to the stripping radius derived by \citet{mccarthy2008}, 
based on more realistic distributions of 
$\rho_{\rm tot}$ and $\rho_{\rm ism}$.

In the above modeling,
the ISM inside $R_{\rm ism}$ is displaced downwards, 
but is still bound to the galaxy,
and keep receiving the ram pressure
and transmitting it via gravity force 
to all the components of the galaxy.
Therefore, this $R_{\rm ism}$ can be identified
with the interaction radius $R_{\rm int}$
in Eq.~(\ref{eq:rp}) and Eq.~(\ref{eq:vis}).
Numerically, we find
\begin{equation}
R_{\rm int} \sim R_{\rm ism} \sim 
6~{\rm kpc} \times
\left( \frac{m_{\rm gal}}{1\times 10^{11}M_\odot} \right)
\left( \frac{v}{10^{3} {\rm km/s}} \right)^{-2}
\left( \frac{n_{\rm ism}}{10^{-2}} \right)
\left( \frac{n_{\rm e}}{10^{-3}} \right)^{-1}
\label{eq:R_ism_num}
\end{equation}
where $n_{\rm ism}$ is the ISM number density 
corresponding to $\rho_{\rm ism}$, 
in units of cm$^{-3}$.
This indeed justifies our assumption of $R_{\rm int} \sim 10$ kpc,
employed when calculating the heating/cooling balance
in Fig.\ref{fig:cool-heat}.
Furthermore, the assumed ISM mass within $R_{\rm ism}$ is 
$M_{\rm ism} = 
\left( \frac{4 \pi}{3} \right) R_{\rm ism}^3\:\rho_{\rm ism}
\sim 1 \times 10^{8}~M_{\odot} \times
(R/6{\rm kpc})^3 (n_{\rm ism}/10^{-2})$.
This is only $\sim 0.1\%$ of the assumed galaxy mass,
and would apply even to elliptical galaxies.
Elliptical galaxies residing in a cluster core region
may have a lower value of $n_{\rm ism}$,
but then the dynamical friction will supersede.
Overall, the ram pressure effect on a member galaxy 
closely resembles the drag force exerted on a smooth solid body
placed in a flowing fluid.

Equation~(\ref{eq:R_ism}) implies
that $R_{\rm ism}$ scales with $m_{\rm gal}$,
as a simple consequence of larger binding energies
in more massive galaxies.
Then, like the dynamical friction case,
this might appear to predict
a strong $m_{\rm gal}$ dependence 
of the cosmological galaxy in-fall, 
and contradict to the results of \citet{gu2013a} and \citet{gu2016}.
However, the estimated $R_{\rm ism}$ is 
already close to the size of galaxies,
so it would not increase very much 
even for more massive galaxies.
Furthermore, galaxies with smaller $m_{\rm gal}$,
mostly spirals, would have higher $n_{\rm ism}$,
which would partially compensate for the lower $m_{\rm gal}$.
Therefore,
we expect relatively weak $m_{\rm gal}$ dependence
of the proposed ICM vs. galaxy interaction mechanism,
in agreement with the actual observations.

While the above estimate assumes hydrodynamical effect only,
in reality we need to take into account the magnetic fields
both in the ICM and the galactic ISM.
For example, MHD simulations of the ram pressure interactions 
by \citet{ruszkowski2014} and \citet{shin2014} 
show a compression of upstream galactic magnetic fields
that are exposed to the incoming ICM flow. 
The consequent smaller gyroradius makes it even harder 
for the ICM particles to flow through the interstellar space.
Furthermore, MFLs in the ICM could be draped 
around each moving galaxy \citep{vi2017},
and effectively increase $R_{\rm int}$.

\subsection{The temperature and thermal stability of the CCC}
\label{subsubsec:RTV}

Although we have so far assumed 
the core-region ICM to be homogeneous,
in reality we need to consider the 2P structure found therein
(\S~\ref{subsec:1st_set}),
consisting of the hot phase and the cool phase (=CCC).
As postulated in our static assumptions,
the two phases are likely to be thermally insulated by MFLs,
and intermixed on spatial scales of $\sim 10$ kpc
in an approximate pressure equilibrium.
These 2P regions of typical clusters 
are still dominates by the hot phase,
and the CCC occupies only a small ($< 20\%$) volume fraction
\citep{ikebe1999,takahashi2009,gu2012}
except the central a few tens kpc 
where the cD galaxy dominates.
Therefore, the heating luminosity $Q_+$  in Fig.~\ref{fig:cool-heat}
must be deposited mainly on the hot phase,
and the generated heat will be carried along MFLs
by the electron conduction and turbulence propagation, 
on time scales of $10^6$ years \citep{gu2012}
and Eq.~(\ref{eq:tau2}), respectively.
As a result, the hot phase will become isothermal
throughout the core region,
as confirmed by observations \citep{ikebe1999,takahashi2009, gu2012}.

Although the electron conduction would not 
work between the two phases,
the ICM turbulence excited in the hot phase 
by galaxies will perturb, from outside, 
the magnetic flux tubes that confine the CCC,
because of Eq.(\ref{eq:beta}) and
Eq.(\ref{eq:4_velocities}).
The turbulence can hence propagate
from the hot phase into the cool phase across the MFLs,
to be dissipated efficiently on the CCC which has a higher density.
In addition, sometimes galaxies would pass through the cool phase, 
and directly excite the turbulence therein
(and also would disturb the magnetospheric structure).
These processes are considered to provide the CCC
with the necessary heating luminosity.

An immediate question would be 
how $Q_+$ is divided into the two phases 
so as to match their respective X-ray luminosities,
and how the 2P configuration is  kept thermally stable
for much longer than the nominal cooling time.
This question  stems from the fact 
that the volume cooling rate of a plasma is 
proportional to $n_{\rm e}^2$ as in Eq.(\ref{eq:Q-}),
whereas the rate of any heating mechanism 
would be proportional to $n_{\rm e}$ (e.g., Eq.\ref{eq:Q+}). 
Then, if the energy deposit on the cool phase is
lower than  its X-ray output luminosity,
the cool phase would lose pressure, 
compressed to become denser, 
and would cool more rapidly.
The cool phase would collapse 
in  $\sim 0.1 t_{\rm H}$ or shorter.
This is virtually identical to the initial cooling-flow problem.

The assumption S3 made in \S~\ref{subsec:1st_set}
helps us to avoid the above difficulty,
because the CCC confined within thin magnetic flux tubes
can be made thermally stable 
(Paper I; \citealt{takahashi2009}; \citealt{gu2013}), 
by so-called RTV mechanism
as quoted in our assumption D4.
Originally proposed for Solar coronae by 
Rosner, Tucker, \& Vaiana (1978),
it was improved by \citet{aschwanden2002},
and confirmed in actual observations
with the Solar Observatory {\it Yohkoh} \citep{kano1996}.
This mechanism assumes 
that a plasma is (1) confined by some external pressure,
(2) within a thin magnetic loop, 
(3) in a hydrostatic equilibrium,
and (4) is heated from outside.
The conditions (1), (2), and (4) are satisfied by our CCC view.
The condition (3) is also likely to hold,
because the filamentary structures in the Perseus 
and Centaurus clusters have constant internal pressures
\citep{fabian2005,sanders2016},
so the momentum flows 
through the filaments are considered rather small.

The RTV mechanism  further postulates
that the plasma within the tube is cooled by
the X-ray radiation,
and by MFL-aligned classical heat conduction 
towards the two footpoints.
Then, if the cooling overwhelms the heating,
the tube becomes thinner due to the reduced internal pressure,
so the conductive flux along the tube decreases.
In addition, a part of the plasma will flow
into the footprints,
to reduce the radiative output.
Thus, the CCC achieves a new equilibrium 
with a reduced cooling luminosity, 
while the plasma temperature remains rather unchanged. 
Conversely, when the heating overcomes the cooling, 
the magnetic tubes will expand,
and additional plasma will be supplied
by the cD galaxy into the tubes.
These lead to a higher radiative and conductive cooling,
which will balance the excess heating. 
The CCC thus responds to 
variations in the heating luminosity,
by adjusting its X-ray radiative output,
rather than changing the temperature.
As a result, the 2T configuration 
keeps the required thermal stability.

In addition to the above evaluation,
we have a more quantitative support indicating 
that the RTV mechanism is actually working in cD clusters.
The RTV theory predicts a scaling law:
the temperature inside the magnetic loops
is determined by the loop half length $H$
and the confining pressure $p_0$ 
(the hot phase pressure in the present case),
without depending on the heating or cooling luminosity
\citep{rosner1978, aschwanden2002}.
Then, according to \citet{takahashi2009} and \citet{gu2012},
the maximum temperature of the plasma confined 
within a cool-core-sized magnetic loop is predicted as
\begin{equation}
T_{\rm c}^{\rm max} = (2.0-2.5) \left[ \left( \frac{p_0}{10^{-10}} \right) 
                     \left( \frac{H}{30\,{\rm kpc}} \right) \right]^{1/3}~~{\rm keV}
\end{equation}
where $p_0$ is in units of dyn cm$^{-2}$.
Considering that this $T_{\rm c}^{\rm max}$ 
refers to the loop-top temperature,
and the volume-averaged loop interior 
must be somewhat cooler \citep{gu2012},
the average CCC temperature is expected to be $1-2$~keV,
in a very good agreement with 
what is observed from typical cool core clusters.  
Further considering the hot phase properties,
\citet{takahashi2009} derived 
a theoretical scaling relation as
$T_{\rm c} \propto T_{\rm h}^{3/4}$
which can approximately reproduce the empirical scaling relation of
 $T_{\rm c} = (0.4-0.6) T_{\rm h}$ 
\citep{allen2001, kaastra2004}.
These results give a strong support to our assumption D4,
because no other convincing explanation has ever been given 
either to the value of $T_{\rm c}$ 
or its dependence on $T_{\rm h}$.

\subsection{Outlook: XRISM and Athena}
\label{subsubsec:outlook}

The X-Ray Imaging and Spectroscopy Mission (XRISM, 
to be launched in 2022), planned as a successor to {\it Hitomi}, 
will carry an X-ray micro-calorimeter
that is nearly identical to the one on {\it Hitomi}. 
Beyond that, we can look forward to Athena \citep[early 2030s]{nandra2013} with even higher spectral resolution and better sensitivity. 
These future X-ray missions will extend the innovative 
{\it Hitomi} results on the Perseus Cluster to many other systems, 
and will provide breakthroughs in our understanding of
the thermal evolution of the ICM. 
As noted in the XRISM mission concept paper \citep{tashiro2018}, 
the scenario of ICM heating from galaxy motion 
becomes one of the feasible alternatives
(but not necessarily mutually exclusive)
to the AGN heating model, to be tested using XRISM. 
Here we describe these future prospects in two relevant aspects.

\begin{figure}[!htbp]
\resizebox{\hsize}{!}{\includegraphics[angle=0]{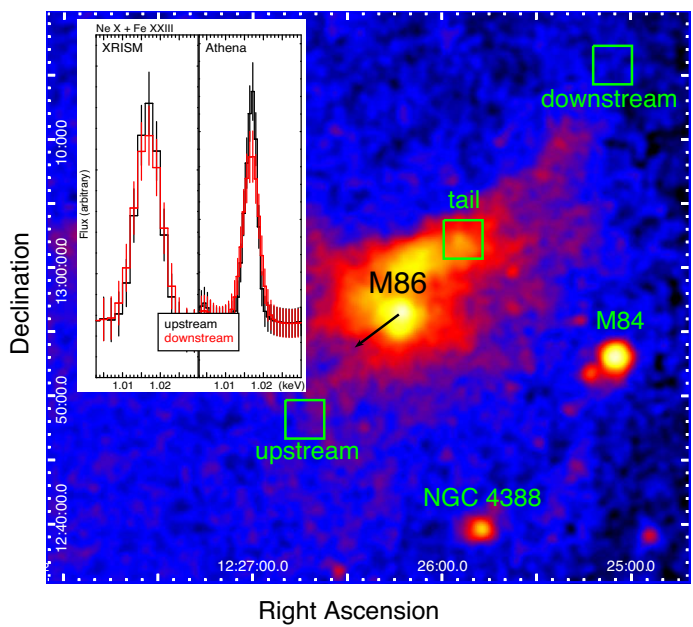}}
\caption{A {\it ROSAT image of the} M~86 region. 
The approximate direction of M~86 motion is marked by a black arrow. 
The green boxes show the field of views of the simulated XRISM/Athena spectra
for the upstream, tail, and downstream regions. Inset: the simulated
XRISM (left, 400~ks) and Athena (right, 30~ks) spectra of a narrow band
around the blended \ion{Ne}{X} and \ion{Fe}{XXIII} lines at 1.017~keV.
The turbulence in the downstream region (red) is assumed to be 565~km s$^{-1}$,
higher than the one in the upstream region (black) by $\Delta \sigma= 400$ km s$^{-1}$.}
\label{fig:m86}
\end{figure}

\subsubsection{Galaxy-ICM coupling}
\label{subsub:coupling}

Our scenario predicts 
that a fast-moving bright galaxy will 
partially drag the local ICM around it.
This prediction can be tested, in principle,
by comparing the local X-ray redshifts around bright member galaxies,
with their optical line-of-sight peculiar motions 
(e.g., \citealt{witt2019}).

One of the best efforts 
to detect the above effect, using the X-ray CCD, 
was made by \citet{tamura2014} for the Perseus cluster. 
By analyzing the well-calibrated {\it Suzaku} Fe-K data, 
the authors determined the radial velocities of the ICM 
in a spatial mesh of $\sim 50-100$~kpc over a
central 600~kpc region, 
with an instrumental systematic uncertainty of $\sim 300$ km s$^{-1}$, 
and compared the ICM motion with the optically measure 
galaxy velocities.
Despite the limited energy resolution of CCD,
their work discovered a mild positive correlation 
between the two components in their line-of-sight motions.
Similarly, a tentative correlation 
has been reported by \citet{yu2016} for A~85. 
The {\it Hitomi} spectrum has the sufficient energy resolution 
to determine the ICM bulk motion 
within a few tens km s$^{-1}$ \citep{hitomi_tur}, 
but the observed region was too small 
to cover a sufficient number of non-cD galaxies in Perseus.

To finally verify the possible galaxy-ICM coupling, 
it is essential to map the ICM motion accurately over a large area. 
We hence simulated a mosaic XRISM observation of the Perseus cluster 
based on the {\it Suzaku} measurement by \citet{tamura2014}. 
The XRISM spectra were simulated for $\sim 80$ pointings,
using the temperature and density profiles 
in \citet{irina2014} and the abundance profile in \citet{werner2013}. 
A systematic uncertainty by 1~eV, arising
from gain calibration limitations, was taken into account.
Even with a short 30~ks exposure for each pointing, 
we confirmed that the XRISM spectrum can determine the 
ICM Doppler redshift within $\sim 50$ km s$^{-1}$, 
up to a large radius of 600~kpc from the cD galaxy. 
By comparing the X-ray data from $\sim 80$ pointings
with the existing optical survey data (e.g., \citealt{huchra1995, huchra2012}),
we expect we can derive a meaningful answer to the prediction.

\subsubsection{Ram pressure}

As described in \S~\ref{sec:turbulence}, 
we expect that the ram pressure and the related MHD effects 
are at least partially responsible for the observed 
turbulence in the cluster central region. 
This scenario is supported by some of the latest simulations 
(e.g., Figure~5 of \citealt{roediger2015} and \citealt{vi2017}), 
which show that both the stripped ISM and downstream ICM 
are perturbed by the ram pressure effect to become more turbulent. 

The micro-calorimeters on-board the upcoming XRISM 
and Athena missions will provide the first direct test 
to the above predictions. 
Here, we show a simulation of the expected results 
for a candidate target, M86, which
is a nearby elliptical galaxy falling into the Virgo cluster 
with a line-of-sight impact velocity of $\sim 1500$ km s$^{-1}$. 
It hosts the brightest ram pressure tail in X-ray and H$\alpha$, 
with a length of $\sim 150$ kpc in projection.
red{High-resolution} X-ray spectra of the upstream, tail, 
and the downstream regions (see Figure~\ref{fig:m86} for their
rough locations) will enable us to search for 
possible turbulence caused by M~86, 
and to constrain its energy injection to the ICM.
The simulation employs the temperature, 
abundances, and emission measures of the M~86 ISM, 
as well as of the surrounding Virgo ICM, 
as measured with {\it Suzaku} \citep{hishi2017}.
Assuming that the turbulence is 165 km s$^{-1}$ \citep{hitomi2016} 
at the upstream, and is boosted by $\Delta \sigma= 400$ km
s$^{-1}$ at the tail and downstream, 
the enhancement would be detected with XRISM,
if we invest rather long exposures,
$\sim 400$~ks at the upstream and downstream, 
and 100~ks at the tail.
With Athena, a much reasonable exposure of $\sim 20-30$~ks 
each region would be sufficient to detect
the turbulence enhancement at the 3$\sigma$ confidence level. 
For lower contrast, e.g., for $\Delta \sigma = 200$ km s$^{-1}$,
a longer exposure will be needed;
1~Ms with XRISM and 60~ks with Athena. 

\section{Summary}


Based on a number of data-oriented X-ray studies, and
by adopting simple assumptions on the static and dynamical aspects, 
we develop a physical model 
which involves the member galaxies to interact with the
cluster environment (particularly the ICM)
through ram pressure, g-mode, and other MHD effects.
As a result, a large amount of energy
is expected to flow from the moving galaxies into the ICM, 
first in a form of turbulence, which then dissipates into heat 
that may significantly offset the long-term 
radiative cooling of the ICM.
The evidence of cosmological infall of member galaxies,
which is a natural consequence of the above view,
has actually been found in previous works.
Our scenario is also well in line with 
the recent {\it Hitomi} result,
because the galaxy-ICM interaction predicts
a mild subsonic turbulence with a good spatial uniformity,
just as observed with  {\it Hitomi}.
We speculate that the galaxy-ICM interaction 
might be universally present across the cluster field, 
creating a significant and persistent energy flow 
in the present Universe.









  


\bibliographystyle{aa}
\bibliography{main}
\end{document}